\documentstyle[psfig]{l-aa} 
\begin{document}
\def\lya{Ly$\alpha$}     %
\def\lyb{Ly$\beta$}      %
\def\Wr{w_{\mathrm r}}   %
\def\Wo{w_{\mathrm obs}} %
\def\mr{m_{\mathrm R}}   %
\def\Mr{M_{\mathrm R}}   %
\def\mb{m_{\mathrm B}}   %
\def\Mb{M_{\mathrm B}}   %
\def\mi{m_{\mathrm I}}   %
\def\Mi{M_{\mathrm I}}   %
\def\mq{m_{\mathrm 450}} %
\def\ms{m_{\mathrm 702}} %
\def\mh{m_{\mathrm 814}} %
\def\kms{~km~s$^{-1}$}   %
\def\cm2{~cm$^{-2}$}     %
\def\za{z_{\mathrm a}}   %
\def\zd{z_{\mathrm d}}   %
\def\ze{z_{\mathrm e}}   %
\def\zg{z_{\mathrm g}}   %
\def\h50{h_{50}^{-1}}    %
\def\hi{H\,{\sc i}}      %
\def\alii{Al\,{\sc ii}}  %
\def\caii{Ca\,{\sc ii}}  %
\def\civ{C\,{\sc iv}}    %
\def\crii{Cr\,{\sc ii}}  %
\def\feii{Fe\,{\sc ii}}  %
\def\mgi{Mg\,{\sc i}}    %
\def\mgii{Mg\,{\sc ii}}  %
\def\ovi{O\,{\sc vi}}    %
\def\znii{Zn\,{\sc ii}}  %
\thesaurus {03(11.17.1; 11.08.1; 11.09.4)} 
\title{The nature of intermediate-redshift damped \lya \ 
absorbers\thanks{Based on observations made with the NASA/ESA {\it Hubble 
Space Telescope}, obtained at the
Space Telescope Science Institute, which is operated by the Association of 
Universities for
Research in Astronomy, Inc., under NASA contract NAS 5-26555}}
\author{V. Le Brun \inst{1}, J. Bergeron \inst{2,3}, P. Boiss\'e \inst{4}, and 
J.M. Deharveng \inst{1}}
\institute{Laboratoire d'Astronomie Spatiale du C.N.R.S., B.P. 8, F-13376 
Marseille CEDEX 12, France, vlebrun@astrsp-mrs.fr, jmd@astrsp-mrs.fr
\and European Southern Observatory, Karl-Schwarzschild-Stra\ss e 2., D-85748
Garching b. M\" unchen, Germany, jbergero@eso.org
\and Institut d'Astrophysique de Paris, CNRS, 98bis boulevard Arago, 
F-75014 Paris, France
\and Ecole Normale Sup\'erieure, 24 rue Lhomond, F-75005 Paris, France, 
boisse@ensapa.ens.fr
}
\offprints{V. Le Brun}
\date {Received May 24th, 1996; accepted Oct 24th, 1996}
\maketitle
\markboth{}{V. Le Brun et al.: The nature of damped \lya \ absorbers}
\begin{abstract}
We present  HST/WFPC2 high-spatial resolution images in the $R$ and $B$ bands
of the close environment of the sightlines to seven quasars which spectra show 
either  a damped \lya \ absorption line,  21~cm absorption, or a very strong 
\mgii /\feii \ absorption system  at intermediate redshifts ($0.4\le z\le 1$). 
Objects  down to about 0.3\arcsec, or 2.0~kpc at $z=0.6$ ($H_0 = 
50$~kms~s$^{-1}$~Mpc$^{-1}$, $q_0=0$), and to a limiting magnitude 
$m_{\mathrm 702, lim}=25.9$ could be detected for seven fields comprising eight 
absorbers (one at higher redshift $z$ = 1.78  towards  MC~1331+170)
 with high \hi \ column densities of at least $1\times 10^{20}$~cm$^{-2}$.

In each case, a candidate absorber with absolute magnitude  
$\Mb \simeq-19.0$ or much 
brighter has been detected. This small sample of gas-rich galaxies at 
intermediate redshifts covers a wide range in morphological types. There are 
three spiral galaxies of various sizes and luminosities (towards 3C~196, 
Q~1209+107 and MC~1331+170), three compact objects (towards EX~0302$-$223, 
PKS~0454+039 and, at high redshift, MC~1331+170), and two amorphous, low 
surface brightness galaxies (towards PKS~1229$-$021 and 3C~286). 
In the fields around 3C~196, PKS 1229$-$021 and Q~1209+107, there is an 
excess of galaxies in the PC2 images, suggestive
of the presence of a group of galaxies associated with the damped \lya \ 
absorber, or maybe with the quasar itself for the two $z_e \leq$ 1.0 
cases. For 3C~196 and 3C~286, the quasar host galaxies have also 
tentatively been discovered. We do not detect any quasar multiple 
images, implying no large amount of dark matter around the damped \lya \ 
absorbers.

This survey also led to the discovery of the first  $z \simeq$ 1.0  
optical counterpart of a quasar radio jet (PKS~1229$-$021).\par
As will be reported elsewhere (Boiss\'e et al. 1996), spectroscopy with the 
HST-FOS of the strong \mgii /\feii \ absorption systems confirms the validity 
of our selection criterion in predicting the existence of damped \lya \ 
systems.
\end{abstract}   
%
\section{\label{intro}Introduction}
Metal-rich absorption line systems (\civ \ and \mgii, damped \lya \  and
21 cm systems) are a very powerful tool for studying the statistical properties
of high-redshift young galaxies, otherwise very difficult to detect
directly, but it must be ascertained which kind of population does indeed probe
each type of system. The large column density absorbers giving rise to 
high-redshift damped
\lya \ absorption lines in quasar spectra  are generally assumed to trace 
proto-galactic disks (Wolfe et al. 1986), whereas \mgii \ absorptions at
intermediate redshifts trace the gaseous halos of luminous field galaxies 
(Bergeron \& Boiss\'e 1991, thereafter BB91; Steidel 1993) with typically 
radii of the order of $75\h50$~kpc (where $h_{50}$ is the Hubble 
constant in units of 50~\kms ~Mpc$^{-1}$, and using $q_0=0$). The main 
arguments  
that have led to associate the  damped \lya \ systems (DLAS) with the 
progenitors of
present-day gas-rich galaxies are related to their mass density (Wolfe 1987) 
and to their metal
content (Pettini et al. 1994). Nevertheless, the population producing  the 
damped \lya \ absorption lines is not yet unambiguously identified since the 
systems 
detected in the optical range are at too high redshifts ($\zd \ge 2$)
to easily detect the absorber by its emission, both because of its apparent 
faintness and  its proximity to the quasar image. 

The aim of this project is to determine the magnitude, morphology, color and 
extent of the \hi \ component of the galaxies causing 21 cm/damped \lya \ 
absorption at intermediate redshift and  investigate whether 
these absorptions probe galactic disks, gaseous halos or elongated whisp-like
structures as in NGC 3067/3C~232 (Carilli \& Van Gorkom 1989). This program is a 
first step towards  relating the properties of present-day
and intermediate-redshift gas-rich galaxies. When the project was undertaken,
 only a few 21 cm absorbers at intermediate redshift were known and no damped 
\lya \ system had yet been discovered at $ 0.2 \le z \le 1.0$ from ultraviolet 
observations. As photoionization modeling had shown that \mgii \ systems with 
very strong \feii \ associated absorption should have \hi \ column densities 
in excess
 of a few 10$^{19}$ cm$^{-2}$ (Bergeron \& Stasi\'nska 1986), which is indeed 
the case for 21~cm absorbers, we also selected absorption systems displaying 
this property.

High spatial resolution images of the selected quasar fields were taken 
with the 
HST-Wide Field and Planetary Camera 2 (WFPC2), 
and UV spectroscopy was obtained with 
the HST-FOS to derive the \hi \ column densities, the gas temperature for the 
21~cm absorbers, and to set constraints on the heavy element abundances. The 
selected quasars exhibit at least one of the following 
properties:
\begin{itemize} 
\item a 21~cm absorption: 3C~196, PKS~1229$-$021 and 3C~286, 
\item a high rest-frame equivalent width ratio $\Wr($\feii$)/\Wr($\mgii$) 
\sim 1$: EX~0302$-$223, Q~1209+107, PKS~0454+039, MC~1331+170,
\item a damped \lya \ line at higher redshift: MC~1331+170, 
\item a galaxy very close to the quasar sightline: PKS~1229$-$021 (Bergeron,  
unpublished CFHT observations), Q~1209+107 (Arnaud et al. 1988), 3C~196 
(Boiss\'e \& Boulade 1990). 
\end{itemize}

Four of these quasar fields have been previously studied with either the HST Wide
Field Camera 2 (3C~196: Cohen et al. 1996) or ground-based telescopes 
(PKS~1229$-$021  
and 3C~286: Steidel et al. 1994a, PKS~0454+039: Steidel et al. 1995). In each 
case, a candidate absorber was detected and these results will be discussed 
and compared to our higher spatial-resolution
observations in Sect.~\ref{indfi}.

In this paper we present the results obtained with the Planetary Camera 2 
(PC2) for seven fields, seven damped \lya \ candidate absorbers and one 
confirmed damped \lya \ system. The observations, 
the method developed for the quasar image subtraction, and the algorithm used 
for the object detection, classification and magnitude estimate are described 
in Sect.~\ref{obsdr}. The individual fields are presented in
Sect.~\ref{indfi}.
The implications of these observations are discussed in Sect.~\ref{disc}.  The
analysis of our spectroscopic data will be presented in Boiss\'e et al. (1996) 
and the Wide Field Camera 2 (WFC2) observations will be discussed in a subsequent 
paper.
\section{\label{obsdr}Observations and data reduction}
\subsection{\label{obs}Presentation of the observations}
All the data were obtained with the Wide Field Planetary Camera 2 (WFPC2),
using the filters F702W and F450W, with central wavelengths 6900~\AA \ and 
4550~\AA \ respectively, except for MC~1331+170. For the latter, the damped 
\lya \ absorber is at higher redshift, and we used the 
filters F702W and  F814W (central wavelength 8300~\AA ). The 
Journal of the observations is given in Table~\ref{obslog}. 
The zero-points of the magnitude scales are not those adopted in the 
HST-STMAG system. They were taken from Whitmore (1995) and Holtzman et al. 
(1995), since the zero-points of the STMAG system are based on a flat 
spectrum, whereas usual visible ground-based photometry is based on the Vega 
spectrum. Adopting a Vega-type spectrum to define the magnitude zero-points 
implies to add $+0.487$, $-0.760$ and $-1.266$ magnitudes to those of 
the STMAG for the F450W, F702W and F814W filters respectively.

\begin{table*}
\caption[]{\label{obslog}Journal of the observations}
\begin{tabular}{lccll|llll}
\hline\noalign{\smallskip}

Object         & \multicolumn{2}{c}{Coordinates (J2000)} & $\ze$ & $\za^a$ (metal-rich
systems) &Date 
& Filter &$n \times \Delta t$ &$m_{\mathrm lim}$\\
               & R.A.           & Dec                    &       &       &     
         & \ \ \ \ (s)        &                  \\
\hline\noalign{\smallskip}
EX~0302$-$223  & 03 04 50.1 &$-$22 11 57 & 1.400 &{\bf 1.0095 (DLAS cand.)}
& 1994, Jun 4  & F450W  & $4 \times 500$ & 25.27 \\
               &            &           &       & 0.4196 (\mgii)
&              & F702W  & $6 \times 600$ & 25.83 \\
PKS~0454+039   & 04 56 47.1 & +04 00 53 & 1.345 & {\bf 0.8596 (DLAS)}
& 1994, Apr 7  & F450W  & $4 \times 500$ & 25.25 \\
               &            &           &       & 0.072 (\mgii)
&              & F702W  & $6 \times 600$ & 25.75 \\
               &            &           &       & 1.0680 (\civ)
&              &        &                &       \\
               &            &           &       & 1.1536 (\mgii/\civ)
&              &        &                &       \\
3C~196         & 08 13 36.0 & +48 13 03 & 0.871 & {\bf 0.437  (DLAS, 21~cm)}
& 1994, Apr 16 & F450W  & $2 \times 1000$& 25.52 \\
               &            &           &       & 0.871 (\mgii) 
& 1995, Apr 16 & F702W  & $4 \times 900$ & 25.97 \\
Q~1209+107     & 12 11 40.6 & +10 30 03 & 2.191 & {\bf 0.6295 (DLAS cand.)} 
& 1994, Nov 11 & F450W  & $2 \times 1000$& 25.65 \\
               &            &           &       & 0.3930 (\mgii)
&              & F702W  & $4 \times 900$ & 25.86 \\
               &            &           &       & 1.8434 (\mgii)
&              &        &                &       \\
PKS~1229$-$021 & 12 32 00.0 &$-$02 24 05& 1.038 & {\bf 0.39498 (DLAS, 21 cm)}
& 1994, May 18 & F450W  & $4 \times 500$ & 25.22 \\
               &            &           &       & 0.7005 (\civ) 
& 1995, May 18 & F702W  & $6 \times 600$ & 25.83 \\
               &            &           &       & 0.7568 (\mgii)
&              &        &                &       \\
3C~286         & 13 31 08.3 & +30 30 32 & 0.849 & {\bf 0.692 (DLAS, 21 cm)} 
& 1994, Nov 8  & F450W  & $2 \times 1000$& 25.66 \\
               &            &           &       &               
& 1995, Nov 8  & F702W  & $4 \times 900$ & 26.20 \\
MC~1331+170    & 13 33 35.8 & +16 49 02 & 2.084 & {\bf 0.7443 (DLAS cand.)}
& 1995, Fev 7  & F702W  & $6 \times 600$ & 26.04 \\
               &            &           &       & {\bf 1.776  (DLAS, 21 cm)}
&              & F814W  & $2 \times 600$ & 25.41 \\
               &            &           &       & 1.3284 (\mgii)     
&              &        & $2 \times 900$ &  \\
               &            &           &       & 1.4462 (\civ)      
&              &        &                &  \\ 
\noalign{\medskip}\hline
\end{tabular}
\smallskip\par
$^{\mathrm a}$ Absorption systems as currently known (see text for references)
\end{table*}
To compare the HST photometric magnitudes to those obtained for ground-based 
observations and to derive the differential magnitude number counts, we 
have used the color equation 
given by Holtzmann et al. (1995) for the transformation from HST to $UBVRI$ 
magnitude systems together with the mean values $B-V = 0.7$ and $V-R = 0.7$,
 corresponding to an intermediate galaxy type at $z\le 1$ (Frei \& Gunn 1994).
The derived average color terms are $B- \mq = 0.15$ and $R-\ms =0.35$. 
For the F814W data, no correction is needed to 
recover the $I$ magnitude,  the correction-term being very small.

Each quasar was located at the center of the PC2 field. 
The PC2 pixel size is 0.046\arcsec, its field of view is $36\arcsec\times 
36\arcsec$ wide, and the spatial resolution is FWHM = 0.103\arcsec, 
0.123\arcsec \ 
in the $x$ and $y$ directions respectively for the raw data, and FWHM =
0.129\arcsec, 0.143\arcsec \ after the $\sigma=0.5$~pixel Gaussian smoothing 
that we used for the presentation of the data. These values are measured on 
the non-saturated
star of the field of Q~1209+107 (object \#3, see Fig.~\ref{q1209field}), and
are consistent with those given in the WFPC2 handbook. For each 
target, several exposures were obtained (at least four for the 
F702W images), to allow a better rejection of the cosmic rays events and to 
minimize saturation of the quasar. In fact, 
the number of cosmic rays is important (about 1300 per PC2 field for a 900~s
exposure), and more than three exposures are necessary to properly remove them.
The hot and corrupted pixels, which appear on all the exposures, have been 
removed using the chart provided by the STScI. The remaining hot
pixels are removed using an adapted median filter. 
\subsection{\label{phot}Detection of the objects and photometry}
\begin{figure}
\centerline{\psfig{figure=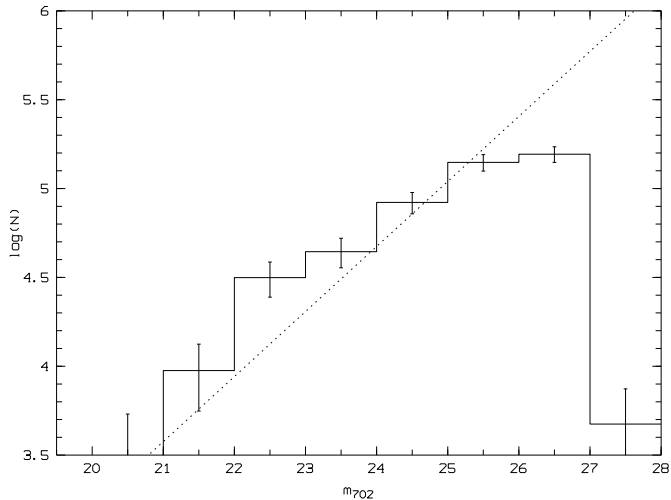,height=7cm,clip=t,angle=-90}}
\caption[]{\label{compt}Differential magnitude number counts in the cumulated 
seven PC2 fields. The dotted line shows the average differential magnitude 
counts from Le Brun et al. (1993). Note the break at the faint end, indicating
the completeness limit, $m_{\mathrm 702, lim} \simeq 26.0$} 
\end{figure}

The detection and classification of the objects present in the PC2 fields were 
made using the software Source Extractor package (Bertin \& Arnouts 1996), 
which offers very robust object detection, high quality deblending, and good 
object classification. 
For the detection, only two parameters have to
be specified: the threshold for the detection and the parameter that controls 
the deblending of the objects, which depends on the
dynamics of the detector. The classification of the objects is quite  
satisfactory, even for the slightly under-sampled images of the WFPC2. 
The  $5\sigma_{\mathrm sky}$ detection threshold, as defined in SExtractor,  
refers to the total flux of an object concentrated and averaged 
over 9 pixels, where $\sigma_{\mathrm sky}$ is the rms 
fluctuation per pixel of the background. 
For the PC2 sampling, this threshold corresponds to a 
$4\sigma_{\mathrm sky}$ detection limit for an unresolved object. For all the 
diffuse extended objects, this selection criterium  always corresponds to a 
 detection limit of at least $2\sigma_{\mathrm sky}$ or 
$\mu_{\mathrm 702, lim}$=22.7 mag arcsec$^{-2}$. The $5\sigma_{\mathrm sky}$ 
detection threshold allows to discriminate between 
very faint stars and cosmic rays events, and to include low surface brightness 
extended objects. This leads to an average limiting magnitude 
$m_{\mathrm 702, lim}=25.9$ for the F702W images 
(see Table~\ref{obslog}). 
Such a value is consistent with the limit of completeness of the fields, as 
illustrated in Fig.~\ref{compt}, where we show the differential magnitude 
number counts of galaxies in the range $20\le \ms \le28$ for the seven 
combined fields. The dotted line indicates the counts of the deep imaging 
survey of Le Brun et al. (1993). The excess in the bins at magnitudes lower 
than 22 are statistical fluctuations due to the low number of galaxies in 
these bins (at most 7), while the excess between $\ms = 22$ and 
$\ms = 24$ 
may be real, and probably due to the presence of some groups near three 
sightlines, 3C~196, PKS 1229$-$021, and, at a lower degree, Q~1209+107 
(see Sects. 3.3, 3.5 and 3.4). However, the bins
between $\ms = 24$ and $\ms = 26$ follow the average distribution, and 
the break occuring in the bin at $\left<\ms\right>=26.5$ indicates 
that the completeness limit of our data is about $m_{\mathrm 702, lim}=26$.
The complete analysis of the galaxy populations of the fields, including 
the WFC2 data, will be made in a subsequent paper.

All apparent magnitudes are given in these filters. For absolute  
magnitudes, $k$-corrections from observed $\ms$ to rest-frame $B$ magnitude 
were estimated using the templates given by Coleman et
al. (1980). We used an Sbc type for all spirals discussed in the paper, and
Im type for the amorphous and/or low-surface brightness galaxies. The
$k$-corrections relative to peculiar objects (e.g the compact absorber 
candidates) are discussed in the text. However, elliptical types have not
been considered, since the presence of large quantities of gas (ascertained
by the DLAS) makes this type unlikely for these objects. 
\subsection{\label{dec}Profile subtraction}
The damped \lya \ absorbers are most probably located very close to the quasar 
sightline. Consequently, it is necessary to search for galaxies superimposed 
onto the quasar image. Since we have observed several 
quasars through the same filter, and roughly at the same location on the PC2 
detector, we have been able to construct an empirical Point Spread Function 
(PSF). For the subtraction to
the quasar image, the latter gives better results than those obtained using a 
library of theoretical PSFs provided by the STScI.

To construct the PSF in an iterative way, we have selected for the first step 
the quasar field for which the candidate absorber is well resolved from the 
quasar image, i.e. Q~1209+107 (see Fig.~\ref{q1209field}). To obtain a
first iteration clean PSF together with a reasonable sampling of the 
background, we have removed the image of the galaxy adjacent on the sky to 
the quasar image and put the corresponding pixels at the average value of the 
surrounding background. This initial PSF has then been used for
subtraction to a second image also barely saturated, i.e
PKS~1229$-$021, after recentering and rescaling. The scaling factor has been 
estimated from the flux ratio in the wings of the quasars, excluding the 
saturated central pixels. As the resulting frame
does not include a strong extended source close to the field center, the 
detected individual objects are clearly resolved from any residual pattern 
due to PSF subtraction. This frame has then been cleaned, all residuals but 
the objects 
being put at the average value of the background (in fact
set to zero for building the PSF frames), and subsequently subtracted from 
the original quasar image to obtain a second PSF frame. The latter is then 
averaged with the initial PSF. The resulting
PSF is then used to analyze the next quasar image.

\begin{figure*}
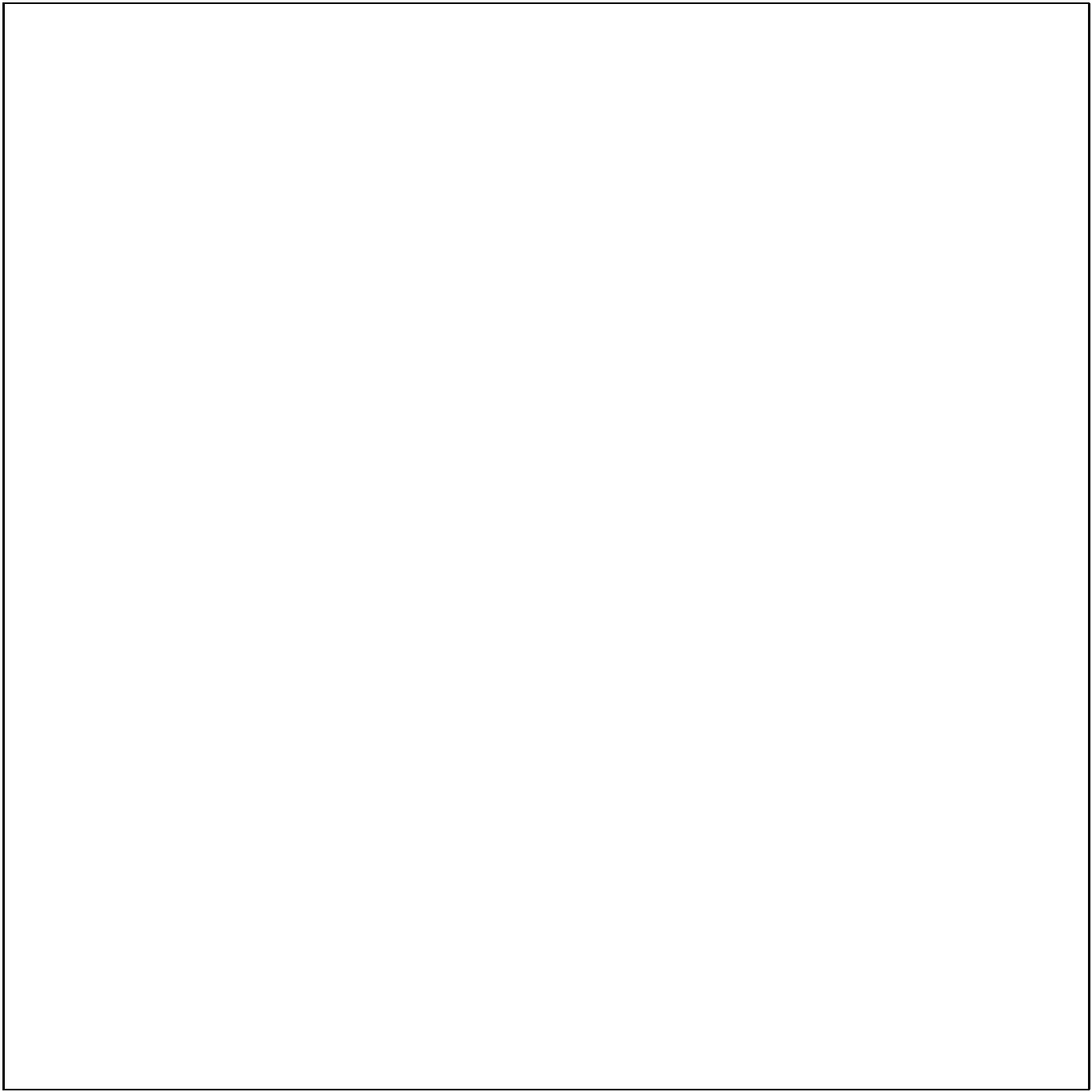

\picplace{18cm}
\caption[]{\label{ex0302field}Field of the PC2 around EX~0302$-$223 after 
smoothing. Objects are labeled as in Table~\ref{ex0302tab}}
\end{figure*}
We thus obtained a PSF built with four point source images of various 
intensities, always located close to the center of the PC2, and therefore 
suffering no differential optical distortions. The images of 3C~196 
and 3C~286 were not used since, in both cases, there is clearly an extended 
underlying object that could not be properly resolved from the PSF residuals. 
The image of PKS~0454+039 was not used either, because the very important 
residuals cannot be clearly ascribed to either the PSF or an underlying 
object, this being most probably due to the high saturation level of the 
quasar. Only the very central parts of the frames obtained after subtraction 
of the final PSF show high residuals due to saturation of the initial 
quasar frames. In each case, we can ascertain the existence of faint objects 
($m_{\mathrm 702, lim} \simeq 25.2$) down to 0.3\arcsec \ from the quasar
center (e.g. Q~1209+107), i.e. $2\h50$~kpc at $z=0.6$, or 0.5\arcsec \ for the 
most saturated quasars (e.g. EX~0302$-$223).
\section{\label{indfi}Description of individual fields}
\subsection{\label{sec0302}EX~0302$-$223 ($\ze = 1.4000$)}
Petitjean \& Bergeron (1990) have detected two metal-rich absorption systems in 
the spectrum of this quasar at $\za = 0.4196$ and $1.0095$. The latter shows
strong \mgii \ and \feii \ absorption lines, and is therefore a possible DLAS. 
In the IUE spectrum of this quasar, Lanzetta et al. (1995) 
have detected \lya \ absorption at $z = 1.0095,~0.9874$ and 0.9690.
The rest-frame equivalent widths $\Wr$ of each of these lines are greater 
than 5\AA, and each could be either damped (N(\hi)$\simeq 1\times 10^{20}$~cm$^{-2}$)
 or multiple \lya \ lines. The 
two lower redshift  systems do not have associated metal absorption lines and 
are thus most likely multiple \lya -only systems, or blends of \lya \ with 
metal lines or Galactic absorptions.  A G270H FOS spectrum of EX~0302$-$223
 will be  retrievable from the HST data base in December 1996 and we will 
then be able to confirm the nature of the $\zd=1.0095$ absorber. 
The  $\za = 0.4196$ \mgii \  absorber has been identified by Guillemin \& 
Bergeron (1996,  hereafter GB), with a $\zg=\za$  very bright galaxy at 
a large impact parameter $D$, with however values consistent with the 
$\left(\Mr,~D\right)$ scaling law given by BB91. A faint galaxy 7.7\arcsec \ 
away from the quasar sightline (object \#7 in Table~\ref{ex0302tab}) has also 
been  identified by GB at $\zg=1.000 \simeq \zd$, but at a too large impact 
parameter ($D=84.3\h50$~kpc) to be the damped \lya \ absorber.  
Finally, a Lyman limit system within $10^4$\kms
\ from the quasar emission redshift has been detected by Koratkar et al. 
(1992) in a IUE spectrum,  and the redshift estimated from the Lyman edge  
 is $\za = 1.33$. 

\begin{figure}
\centerline{\psfig{figure=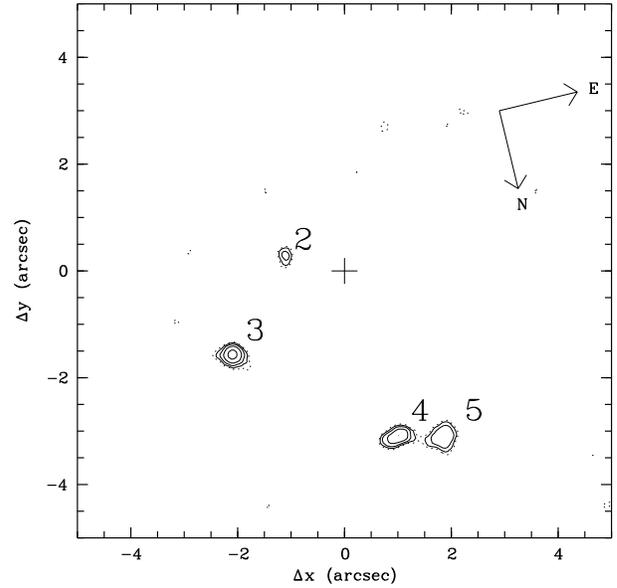,height=8cm,clip=t}}
\caption[]{\label{ex0302sub}Close vicinity of EX~0302$-$223 after PSF 
subtraction. The location of the quasar is marked with a cross. Contour level 
values are 1.5 (dotted), 2, 3, 5 and $9\sigma$ above the mean sky level}
\end{figure}

\begin{table}
\caption[]{\label{ex0302tab} List of the objects in the field of EX~0302$-$223}
\begin{tabular}{rrrrrr}
\hline\noalign{\smallskip}
 Obj. & $\Delta \alpha$ & $\Delta \delta$ & $\theta$ &$\ms$ &$\mq$ \\
      & \arcsec         & \arcsec         & \arcsec  &      &      \\
\hline\noalign{\smallskip}  
  1 &  0.00 &  0.00 &  0.00 &  17.0$^{\mathrm a}$ &  16.55 \\
  2 & -0.98 & -0.57 &  1.14 &  25.4  &     - \\
  3 & -2.24 &  1.04 &  2.47 &  23.8  & 25.1   \\
  4 &  0.30 &  3.21 &  3.22 &  24.0  & 24.5   \\
  5 &  1.04 &  3.41 &  3.57 &  24.2  &  -     \\
  6 &  1.69 & -5.47 &  5.73 &  23.63 &  -     \\
  7 & -2.65 & -7.25 &  7.72 &  22.37 &  -     \\
  8 & -3.56 &  6.97 &  7.83 &  25.38 &  -     \\
  9 & -8.36 & -2.14 &  8.63 &  24.52 &  -     \\
 10 &  4.72 & 11.20 & 12.15 &  25.24 &  -     \\
 11 &  7.52 & 12.11 & 14.26 &  25.05 &  -     \\
 12 &  4.91 &-13.56 & 14.42 &  24.72 &  -     \\
 13 & -8.49 & 11.94 & 14.66 &  21.18 & 24.13  \\
 14 & 15.04 & -0.68 & 15.06 &  18.36 & 19.71  \\
 15 & -8.21 &-12.63 & 15.06 &  25.07 &  -     \\
 16 &-14.92 &  3.04 & 15.23 &  24.56 &  -     \\
 17 &  8.85 & 12.46 & 15.28 &  25.37 &  -     \\
 18 & -0.11 &-15.35 & 15.35 &  25.11 &  -     \\
 19 & -5.99 & 15.47 & 16.59 &  25.45 &  -     \\
 20 & 11.91 &-11.55 & 16.59 &  25.26 &  -     \\
 21 &-16.35 & -2.94 & 16.61 &  25.22 &  -     \\
 22 &-14.89 &  8.82 & 17.30 &  22.53 & 23.25  \\
 23 &-13.54 &-10.97 & 17.43 &  25.06 &  -     \\
 24 & -7.40 & 15.88 & 17.52 &  23.41 &  -     \\
 25 & -9.65 & 14.87 & 17.73 &  20.52 & 24.25  \\
 26 & 15.50 & -9.53 & 18.20 &  24.58 &  -     \\
 27 &-12.34 &-13.87 & 18.57 &  22.39 & 24.13  \\
 28 & -7.40 &-17.56 & 19.06 &  25.44 &  -     \\
\noalign{\medskip}\hline
\end{tabular}
\smallskip\par
$^{\mathrm a}$ Saturated
\end{table}
The whole PC2 field is presented in Fig.~\ref{ex0302field} and  
Fig.~\ref{ex0302sub} shows the 10\arcsec \ square field centered on the quasar 
after PSF subtraction and a Gaussian smoothing with $\sigma =0.5$~pixel. There 
are four faint objects at impact parameters less than 5\arcsec\ (objects \#2 
to \#5), the closest 
being detected only after profile subtraction. These objects have impact 
parameters smaller than 24 and $39\h50$~kpc at $z=0.4196$ and 1.0095 
respectively. 
It is then unlikely that any of them is associated with the lower 
redshift \mgii \ absorber, since large 
\hi \ column densities, thus  strong \feii \ absorption,  are expected  at 
such small impact parameters, but no \feii \ absorption line is  
detected at $z=0.4196$ (Bergeron unpublished).  Furthermore, there is no strong 
emission line at the wavelength of the expected [O\,{\sc ii}]$\lambda 3727$  
line at $z=1.0095$ in the spectra of objects \#4 and \#5, unresolved in 
ground-based observations (GB). 

Consequently, objects \#2 and \#3 are the most likely damped \lya \ absorber
candidates, they
have colors $\mq-\ms>-0.1$ and $\mq-\ms = 1.3$ respectively. Since this does 
not  
constrain their spectral types, we will use the $k$-correction of an Sbc
galaxy. We then obtain $\Mb = -20.4$ and $-22.0$ for objects \#2 and \#3, with
uncertainties as large as 0.5 mag.
Object \#7 would then be another field galaxy at a redshift similar to that 
of the damped \lya\ absorber, and might contribute to the detected \mgii \ 
absorption, which is a triple system spanning 170\kms  (Petitjean \& Bergeron 
1990). Its absolute magnitude is $\Mb=-23.4\pm 0.5$. 

The galaxies \#4 and \#5 are not embedded in a lower surface brightness 
envelope (at the $1\sigma$ rms level above the background), which does 
not support the assumption of a physical pair. They could be
associated with the strong \lya \ absorbers at $\za=0.9874$ and 0.9690.

There are 28 detected objects in this PC2 field brighter than the $5\sigma$ 
threshold or $m_{\mathrm 702, lim}=25.8$. They are listed in Table~\ref{ex0302tab}. 
Object \#14 is a bright, $\Mb=-20.0$, fairly blue (B$-$R=1.3) 
spiral galaxy at $z = 0.118$ (Bergeron, unpublished), which should have 
associated \mgii \ absorption. This expected absorption is within the 
 wavelength range of the FOS-G270H spectrum, not yet available in the HST
archive data base. The other brighter objects in the field, \#13 and \#25, 
have very red colors, $\mq-\ms=2.9$ and 3.7 respectively. They could be 
elliptical 
galaxies at $z \simeq 0.5$ (Frei \& Gunn 1994),  with
impact parameters of about 110-140$\h50$~kpc and might have associated \lya 
-only absorption. The $z=0.4196$ absorber is thus most probably the bright 
galaxy identified by GB at a large impact parameter, outside the PC2 field. 
\subsection{\label{sec0454}PKS~0454+039 ($\ze = 1.345$)}
\begin{figure*}
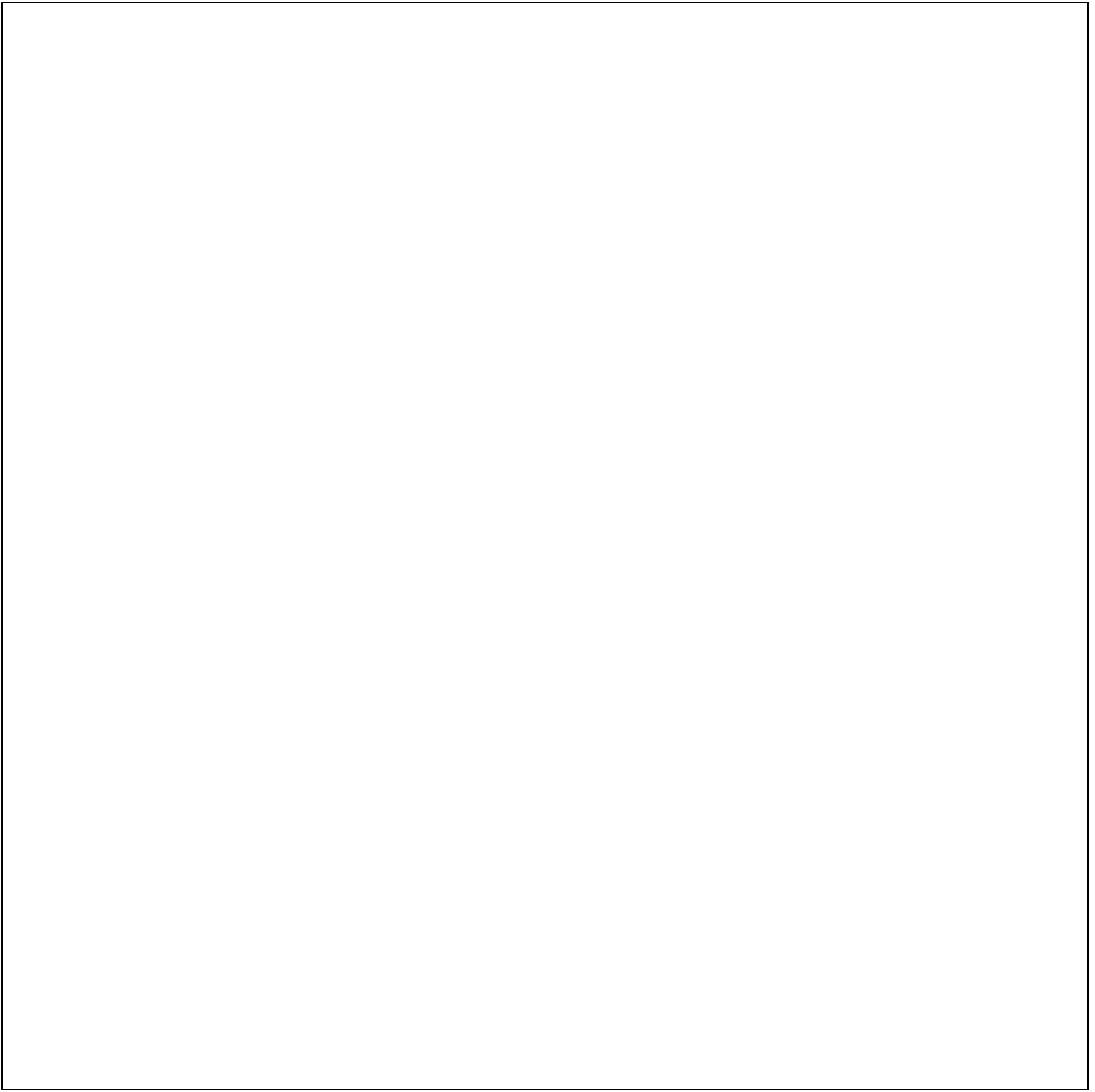

\picplace{18cm}
\caption[]{\label{q0454field}Field of the PC2 around PKS~0454+039. Objects
are labeled as in Table~\ref{q0454tab}}
\end{figure*}
There are two previously known absorption systems in the spectrum of this 
quasar. The $\zd=0.8596$ system has very strong \mgii \ and \feii \ absorption 
lines, which suggests a large $N$(\hi) value. Steidel et al. (1995) have
detected associated \znii \ and \crii \ absorption lines, which are presently 
found only in DLAS. They have also observed the quasar with the HST-FOS, and
confirmed that the \lya \ line of this system is damped 
with N(\hi)= $5.7\times 10^{20}$~cm$^{-2}$. The $\za =1.1536$ 
system shows strong \mgii \ and \civ \ absorption, but no \feii \ absorption 
(Steidel \& Sargent 1992), and the \lya \ line is clearly not damped (Steidel 
et al. 1995). Recently, Boiss\'e et al. (1996) have detected two 
other high-redshift systems: a \civ \  system at $\za = 1.0680$ and a very 
strong \lya -only system  ($\Wr = 3.2$~\AA) at $\za=0.9781$ with 8 associated 
lines from  the Lyman series.

After PSF subtraction performed on 
ground-based images, Steidel et al. (1995) have detected a galaxy 2.1\arcsec \
away from the quasar, i.e $21.7\h50$~kpc at $\zd=0.8596$, that they identified 
as the damped Ly$\alpha$ absorber.
There is a bright object 4\arcsec \ east of the quasar image which is a dwarf
galaxy at  $\zg=0.072$  (Steidel et al. 1993). When the color-term corrections 
are taken into account, our measured magnitudes for 
this galaxy (object \#3) are in good agreement with theirs. 
 No associated \mgii \ absorption
has been reported by Steidel et al. (1995), but a \mgii \  doublet is clearly
present at this redshift in the FOS-G270H spectrum obtained by Boiss\'e et al. 
(1996). There is no object at small impact parameter which could give rise to
the  $\za=1.1536$ absorption system. 
\begin{table}
\caption[]{\label{q0454tab}List of the objects in the field of PKS~0454+039}
\begin{tabular}{rrrrrr}
\hline\noalign{\smallskip}
Object & $\Delta\alpha$  & $\Delta\delta$  & $\theta$ & $\ms$ &$\mq$ \\
       & (\arcsec)       & (\arcsec)       & (\arcsec)&       &     \\
\hline\noalign{\smallskip}
  1 &  0.00 &  0.00 &  0.00 & 16.61 & 17.64  \\
  2 &  0.4  &  0.7  &  0.8  & 24.2  & 25.3  \\
  3 &  3.79 &  1.43 &  4.05 & 20.31 & 21.55  \\
  4 &  4.29 &  4.79 &  6.43 & 25.25 &   -    \\
  5 &  0.15 & -7.51 &  7.51 & 23.61 &   -    \\
  6 &  7.13 &  2.92 &  7.70 & 24.08 &   -    \\
  7 &  6.51 & -5.82 &  8.74 & 23.05 & 24.90  \\
  8 &  9.29 &  0.57 &  9.31 & 22.43 & 23.82  \\
  9 & -8.90 &  7.87 & 11.88 & 24.83 &   -    \\
 10 &-11.35 & -4.24 & 12.11 & 23.89 & 24.63  \\
 11 & 11.71 &  3.88 & 12.34 & 22.61 &   -    \\
 12 & 12.08 & -7.74 & 14.35 & 23.55 &   -    \\
 13 &-10.11 &-10.33 & 14.45 & 25.18 &   -    \\
 14 & -0.98 &-15.57 & 15.60 & 24.32 &   -    \\
 15 &-12.29 & -9.90 & 15.78 & 19.63 & 22.65  \\
 16 & -0.46 &-16.12 & 16.12 & 25.57 &   -    \\
 17 & -9.97 &-15.99 & 18.84 & 23.62 &   -    \\
 18 &-18.35 & -5.29 & 19.10 & 22.60 & 24.19  \\
\noalign{\medskip}\hline
\end{tabular}
\end{table}
\begin{figure}
\centerline{\psfig{figure=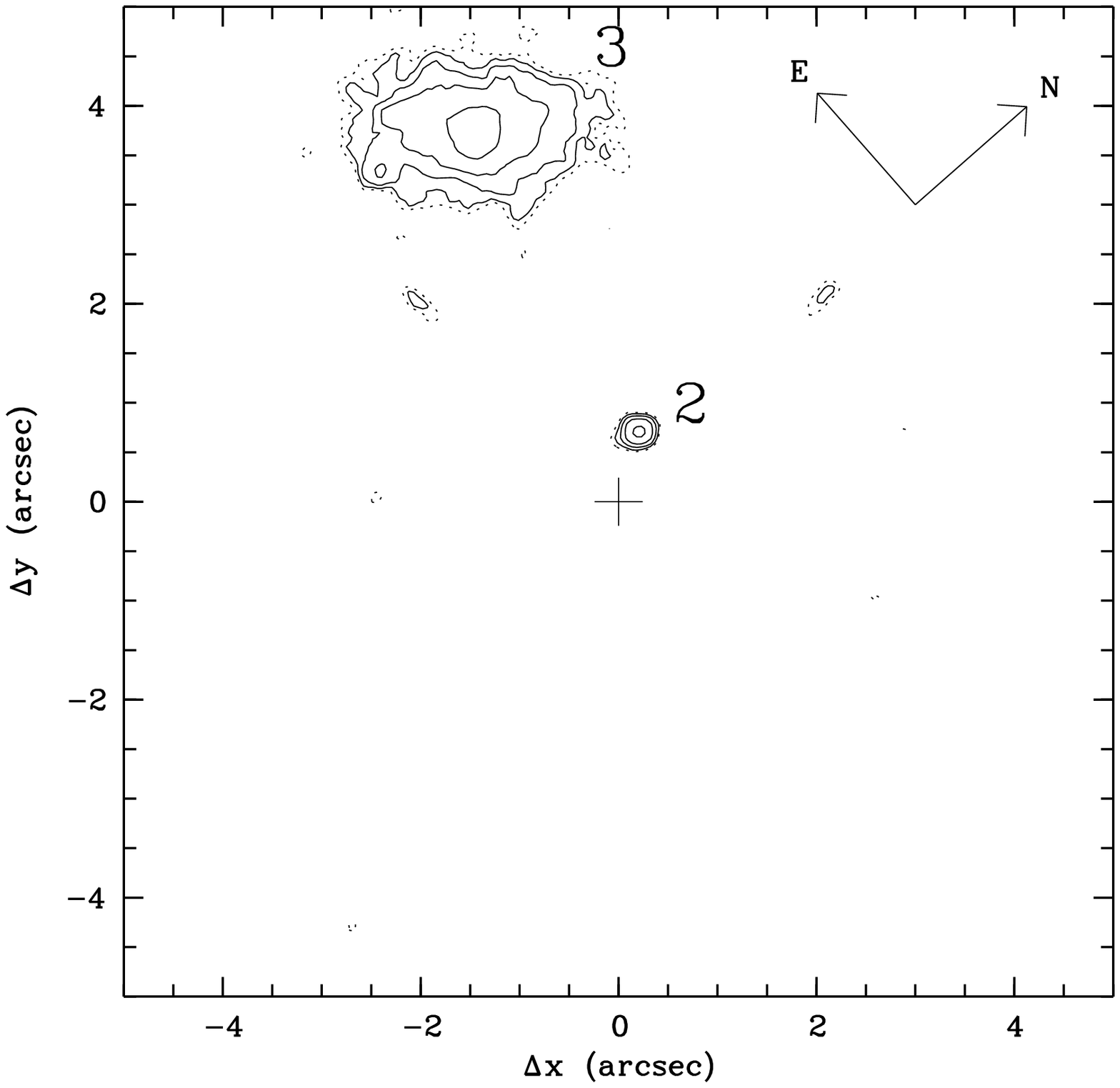,height=8cm}}
\caption[]{\label{q0454sub} Same as Fig.~\ref{ex0302sub} for PKS~0454+039}
\end{figure}
The whole PC2 image is presented in Fig.~\ref{q0454field} and the central part 
of the field is shown in Fig.~\ref{q0454sub} after PSF subtraction and
smoothing. The properties of the galaxies detected in the whole PC2 field are 
listed in Table~\ref{q0454tab}.  The faint galaxy reported by Steidel et al. 
(1995) is just at the edge of the PC2 quasar image and very clearly detected 
after PSF subtraction (object \#2 in Fig.~\ref{q0454sub}). The
impact parameter is smaller by over a factor of two than that measured by 
Steidel et al. (1995), most probably due to a less accurate PSF subtraction 
for the ground-based images; our measured $R$ band magnitude is very close 
to their. Assuming this galaxy to be the damped \lya \ absorber
leads to a small impact parameter, $D=8.3\h50$~kpc. This object is barely
detected in the F450W image, near to the limiting magnitude ($\mq=25.3$), and,
as in the case of objects \#2 and \#3 in the field of EX~0302$-$223, we do
not have any information on the spectral type of this object. Again, we use
a Sbc spectrum to derive $\Mb  = -20.5\pm0.5$. 

There are two galaxies in the field that could be responsible for 
the $\za=1.1536$ \mgii/\civ \ system: objects \#4 and \#5, which would 
then respectively have $\Mb=-20.6$, $D=75\h50$~kpc and $\Mb=-22.2$,
$D=86\h50$~kpc. Object \#5 is a very diffuse, irregular, low surface 
brightness galaxy with a peak intensity $\mu_{\mathrm 702,  max} = 
22.7$~mag~arcsec$^{-2}$. The impact parameter of object \#4 is somewhat larger 
(by $\sim 50\%$) than the value expected for its
magnitude from the $\left(\Mr,D\right)$ scaling law (BB91, GB), whereas for 
object \#5 it is roughly equal to the maximum value given by this scaling law. 
Object \#5 is thus tentatively identified as the  $\za = 1.1536$ absorber. 

Among the 18 PC2 field objects brighter than $m_{\mathrm 702, lim}=25.8$, four are 
classified as stars, but one of them (object \#7) shows some diffuse extension 
suggesting a spiral arm. It is not clear whether it is a galaxy with an 
unresolved nucleus or a foreground star coincident with a background galaxy. 
There are two faint, very diffuse and extended galaxies, objects \#5 
(mentioned above) and \#10 ($\mu_{\mathrm 702,  max} = 22.4$~mag~arcsec$^{-2}$).
\subsection{\label{sec196}3C~196 ($\ze = 0.871$)}
Brown \& Mitchell (1983) found a 21~cm absorption system in 3C~196 at
$\zd = 0.437$. The associated metal lines have been identified by Foltz et 
al. (1988) and Boiss\'e \& Boulade (1990). Five very strong \feii \ 
lines are detected as well as the  Mn\,{\sc ii} triplet and the Ca\,{\sc ii}
doublet which strongly suggests a DLAS. 
Since the radio emission comes predominantly from extended radio lobes, the 
absorber must cover both the optical
quasar and part of the radio lobes, but not the more 
compact radio hot spot located in the northern lobe (Brown et al. 1988), and
the intervening \hi \ absorbing cloud must be larger than $4.5\h50$~kpc. 
Finally, Oren \& Wolfe (1995) have detected a substantial 
Faraday rotation residual toward this radio source, with 
$RRM=-121.8$~rad~m$^{-2}$, even larger than the 
value derived toward the radio-jet of PKS~1229$-$021 by Kronberg et al. 
(1992). Oren \& Wolfe (1995) propose that this Faraday rotation is induced 
by the 21 cm absorber.  This quasar exhibits a second  
\mgii \ system at the quasar redshift ($\za = 0.871$). 

\begin{table}
\caption[]{\label{q3c196tab}List of the objects in the field of 3C~196}
\begin{tabular}{rrrrrr}
\hline\noalign{\smallskip}
Object & $\Delta\alpha$  & $\Delta\delta$  & $\theta$ &$\ms$ &$\mq$ \\
       & (\arcsec)       & (\arcsec)       & (\arcsec)&      &      \\
\hline\noalign{\smallskip}
  1 &  0.00 &  0.00 &  0.00 &  17.83 & 18.71 \\
  2 &$<0.3$ &$<0.3$ &$<0.3$ &$<21.25$ &        \\
  3 & -0.3  &  1.1  &  1.1  &  22.3  &      \\
  4 &  1.21 & -0.86 &  1.48 &  19.96 & 22.23 \\
  5 &  1.19 &  2.30 &  2.59 &  24.81 &   -     \\
  6 &  2.42 &  6.45 &  6.89 &  22.86 & 24.75   \\
  7 &  1.97 & -7.05 &  7.32 &  25.27 &   -     \\
  8 & -7.48 & -1.14 &  7.57 &  23.49 &   -     \\
  9 & -7.12 &  3.00 &  7.73 &  22.04 &   -     \\
 10 & -8.83 &  2.31 &  9.13 &  24.65  & 25.15   \\
 11 &  8.91 &  3.37 &  9.53 &  24.07 &   -     \\
 12 &  9.26 &  4.83 & 10.44 &  21.73 & 24.88  \\
 13 &  5.33 & 10.49 & 11.77 &  25.69 &   -     \\
 14 &  9.55 &  6.98 & 11.83 &  25.47 &   -     \\
 15 &  9.87 &  6.73 & 11.95 &  25.28 &   -     \\
 16 & -1.65 &-12.05 & 12.04 &  22.94 &   -     \\
 17 &-13.16 & -0.05 & 13.16 &  23.91 &   -     \\
 18 & 10.55 & -8.59 & 13.61 &  23.24 &   -    \\
 19 & 10.30 &  8.99 & 13.67 &  24.53 & 25.25  \\
 20 &-12.47 &  7.04 & 14.32 &  23.21 &   -    \\
 21 & -9.36 &-11.02 & 14.46 &  23.95 &   -     \\
 22 & -1.75 &-15.11 & 15.21 &  24.45 &   -     \\
 23 & 13.67 & -8.96 & 16.35 &  22.79 &   -     \\
 24 &  4.21 & 16.22 & 16.76 &  22.89 &   -    \\
 25 & -4.22 &-17.25 & 17.76 &  23.52 &   -    \\
 26 &-18.42 &  2.59 & 18.60 &  23.84 & 25.65   \\
 27 & 18.65 & -4.72 & 19.24 &  24.08 &   -     \\
 28 &-17.90 &  8.99 & 20.03 &  22.38 &   -     \\
 29 &-20.05 &  1.89 & 20.14 &  25.11 &   -     \\
 30 & -6.79 &-20.01 & 21.14 &  24.61 &   -     \\
\noalign{\medskip}\hline
\end{tabular}
\end{table}
On high-spatial resolution images taken at the CFHT, Boiss\'e \& Boulade 
(1990) have detected two galaxies close to the quasar sightline 
(respectively at 1.2\arcsec \ and 1.7\arcsec ). 
They associated the brightest one to the 21~cm absorber, implying
an impact parameter and absolute magnitude of $D=12\h50$~kpc and 
$\Mr = -21.8$ (without $k$-correction). Cohen et al. (1996) have recently 
observed 3C~196 with the HST-FOS (G160L grism) and the WFC2
(pixel size of 0.0966\arcsec, resolution of 0.15\arcsec).  Unfortunately, 
the \lya\ line at $\zd = 0.437$ coincides with  the Lyman limit system at  
$\za \simeq \ze = 0.871$ and these authors could not conclude whether 
the   $\zd = 0.437$ \lya\ line is damped (with 
N(\hi) $\simeq 1.5\times 10^{20}$~cm$^{-2}$) or not. The latter alternative 
would  be inconsistent with the presence of strong associated \feii, 
Mn\,{\sc ii} and  Ca\,{\sc ii} absorption.  
Their $R$-band images reveal that the brightest object is 
a barred spiral galaxy (type SBc) with very extended arms which  cover the 
radio lobes (see their Fig.~1). As  Boiss\'e \& Boulade (1990), they 
conclude that this object could be identified as the 21 cm/\lya\  absorber, 
if a high column density gaseous disk is associated with this luminous 
spiral galaxy. In their PSF subtracted image (using a synthetic PSF generated 
by the Tiny Tim software) they also detect the  northern, fainter object 
(\#3 in Table~\ref{q3c196tab}) close to the quasar image previously 
reported by Boiss\'e \& Boulade 
(1990), which in the WFC2 image extends to approximately 1.2\arcsec \ south of
the quasar. If this galaxy is at $\zg \simeq \ze = 0.871$, as suggested by  
Boiss\'e \& Boulade (1990), they conclude that its proximity to the quasar
together with a lack of associated damped \lya \  absorption would imply that
either this object is not a galaxy or it is affected by the quasar ionizing
radiation field. 
\begin{figure*}
\picplace{18cm}
\caption[]{\label{q3c196field}Field of the PC2 around 3C~196. Objects
are labeled as in Table~\ref{q3c196tab}}
\end{figure*}
\begin{figure}[!h]
\centerline{\psfig{figure=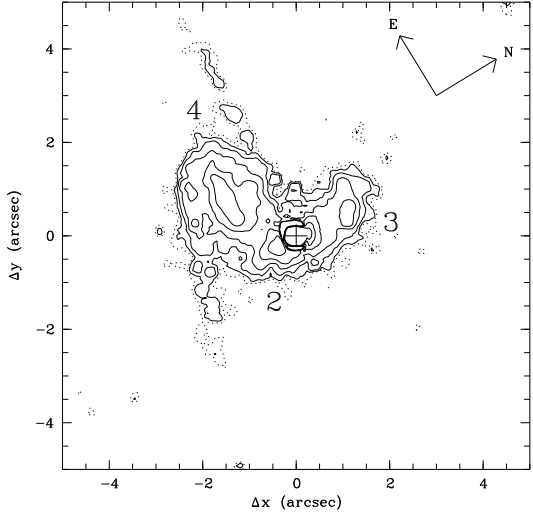,height=7cm}}
\caption[]{\label{q3c196sub}Same as Fig.~\ref{ex0302sub} for 3C~196, with 
levels at 1.5 (dotted), 2, 3, 5, 9 and 17 and $33\sigma$}
\end{figure}
Their analysis of the $\za \simeq \ze$ system led them to suggest that 
the high ionization component of this system may cover only partially the 
quasar emission line region, and its physical properties could then be similar 
to those in BAL quasars. 

Our PC2 image has a resolution about twice as high as that obtained by Cohen 
et al. (1996) and an exposure time in the F702W filter three times longer.
We also have information on the $\mq-\ms$ color for some objects in the field, 
including the bar of the luminous spiral galaxy. The red image is shown in 
Fig.~\ref{q3c196field} and the PSF subtracted central part is given
in  Fig.~\ref{q3c196sub} after weak smoothing. The diffuse northern object 
(\# 3 at $\theta=1.1\arcsec$) is clearly resolved from the quasar, whereas 
the southern object, reported by Cohen et al. (1996), is compact and still
blended with the quasar residual image (object \# 2 at $\theta \leq 0.3 
\arcsec$). This compact object is resolved in the north-south direction 
(i.e. towards the radio hot spot), but not in the east-west one. Its linear 
size along the major axis is roughly 1.5\arcsec \ or $16\h50$~kpc at 
$\zg=0.871$. This object could be related to the host galaxy of the quasar, 
and would then be very luminous, $\Mb<-23.1$. However, this estimate is 
uncertain because the outermost parts of this object are blended on the 
southern side with the barred spiral galaxy and on the northern side with the 
diffuse northern object (\#3).  The continuum emission of the latter amorphous 
galaxy barely covers the northern hot spot, and not the diffuse 
part of the radio lobe. Object \#3 could be associated with the quasar host 
galaxy (e.g. a tidal tail), or be  a galaxy belonging to the group possibly 
associated with the quasar (see below). At $\zg=0.871$, its absolute 
magnitude and impact parameter are $\Mb=-22.0$ and $D=11.5\h50$~kpc. 
The interstellar medium of the galaxies \#2 and \#3 would then be ionized by 
the quasar UV radiation flux for gas densities lower than 10~cm$^{-3}$
and  radial distances up to $25\h50$~kpc. 

The $\mq-\ms$ color of the luminous spiral galaxy equals 2.3 and for the bar 
only $\mq-\ms=2.0$. This color is that of a Sbc galaxy at redshift $z\ga 0.4$
(Frei \& Gunn 1994), which is consistent with this object being the 21 cm
 absorber. Its arms extend over 9\arcsec \ from end 
to end, or about $70\h50$~kpc at $\zd=0.437$. The properties of this galaxy, 
very large extent and high luminosity, $\Mb = -22.1$, are fairly
extreme for its class. In order to study its continuum spectrum and ascertain 
its redshift, one of us (J. Bergeron) has recently observed this object 
with the ARGUS integral field spectrograph at the CFHT. Data reduction 
are in progress, but a quick look at the time of the observing run did not
reveal strong [O\,{\sc ii}]$\lambda 3727$ or H$\alpha$ emission at 
$\zd=0.437$.

In the PC2 field, there are 30 objects brighter than $m_{\mathrm 702,  lim}=
26.0$ (see Table~\ref{q3c196tab}).  As compared to the other
PC2 fields, there appears to be an excess of galaxies in the magnitude range 
$22.0 \le \ms \le 24.0$, statistically significant at the $4\sigma$ level. 
The average number of galaxies in this magnitude range is $4.2 \pm 2.2$ for 
all fields, excluding those of 3C~196 and PKS~1229$-$021, whereas it 
reaches 12 and 9 in these two latter cases (the field 
around PKS~1229$-$021 is discussed in Sect.~\ref{sec1229}). Since radio-loud
quasars are known to lie preferentially in dense environments, this excess of 
galaxies 
most probably traces a galaxy cluster or group associated with 3C~196. 
 The average absolute magnitude of the cluster galaxies detected in the 
PC2 field is $\Mb \simeq -21.4$. A preliminary analysis of the WFC2 adjacent
fields gives a similar $4\sigma$ excess as compared to the expected average 
number density of galaxies per magnitude and per square degree (see e.g.
Le Brun et al. 1993) over a similar range of apparent magnitudes, 
$22.0 \le\ms\le 24.5$, or $-23.6\le\Mb\le-20.6$ at $\ze=0.871$. 
\begin{figure*}[!ht]
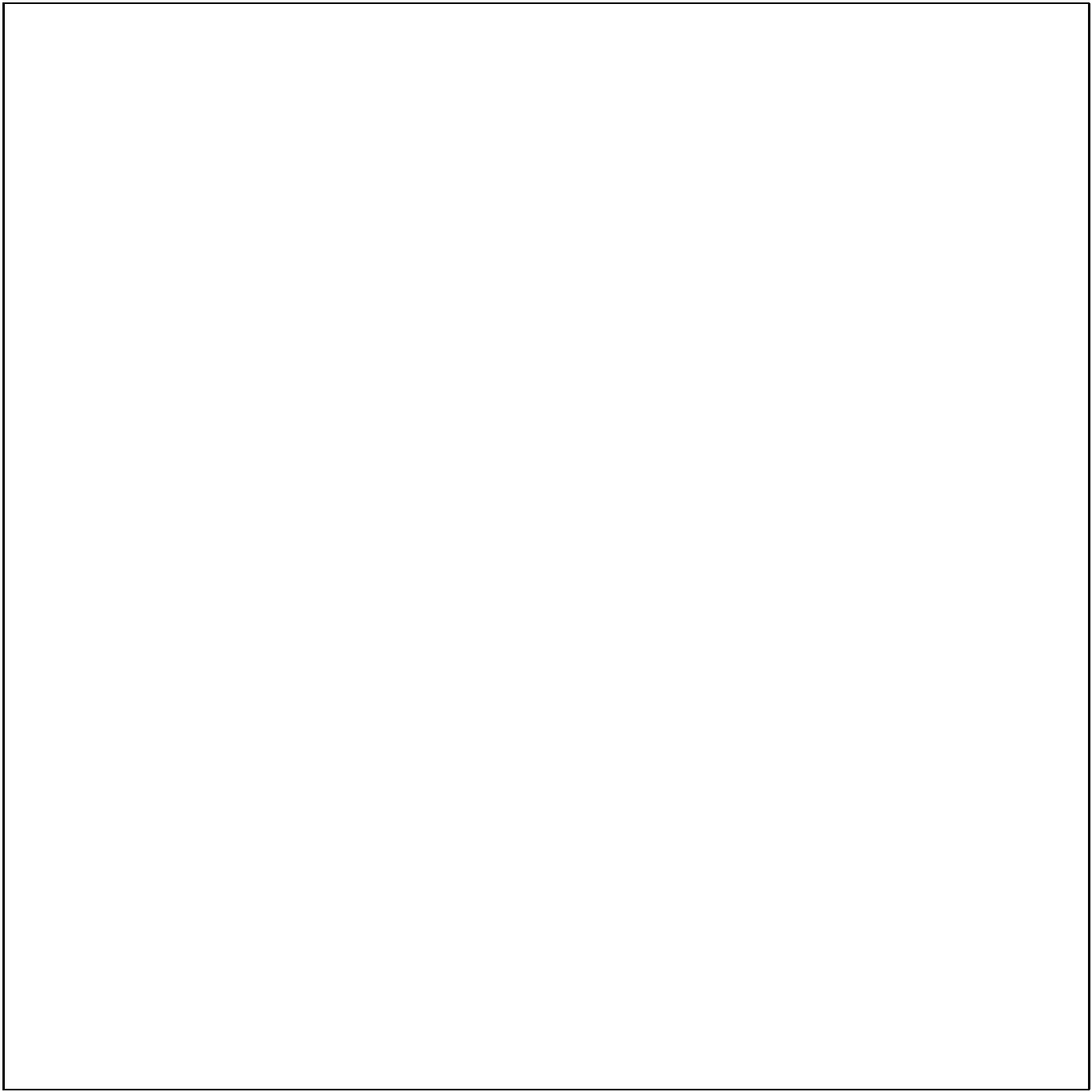

\picplace{18cm}
\caption[]{\label{q1209field}Field of the PC2 around Q~1209+107. Objects
are labeled as in Table~\ref{q1209tab}}
\end{figure*}
\subsection{\label{sec1209}Q~1209+107 ($\ze = 2.191$)}
This quasar exhibits three absorption systems  at $\za = 0.3930$ (\mgii ), 
0.6295 (strong \mgii \ and \feii) and 1.8434 
(\mgii ) (Young et al. 1982). A galaxy at an angular impact parameter 
$\theta = 7.1\arcsec$ and with a magnitude  $\mr = 21.9$ has been
identified by Cristiani (1987) as the $\za = 0.3930$ absorber. In high spatial 
resolution CFHT images, Arnaud et al. (1989) have detected a galaxy at 
1.3\arcsec \ from the quasar sightline, which corresponds to $11.8\h50$~kpc at 
$\zd = 0.6295$. 
\begin{table}[!h]
\caption[]{\label{q1209tab}List of the objects in the field of Q~1209+107}
\begin{tabular}{rrrrrr}
\hline\noalign{\smallskip}
Object & $\Delta\alpha$  & $\Delta\delta$  & $\theta$ &$\ms$ &$\mq$ \\
       & (\arcsec)       & (\arcsec)       & (\arcsec)&      &      \\
\hline\noalign{\smallskip}
  1 &  0.00 &  0.00 &   0.00 &  18.06 & 18.63  \\
  2 &  1.62 & -0.10 &   1.62 &  21.59 & 23.07 \\
  3 & -5.34 &  1.03 &   5.43 &  20.10 & 23.68 \\
  4 & -5.04 &  2.65 &   5.70 &  23.89 &   -     \\
  5 &  5.89 & -3.34 &   6.77 &  25.37 &   -     \\
  6 & -6.66 & -1.87 &   6.91 &  25.00 &   -     \\
  7 & -3.93 & -5.77 &   6.98 &  23.53 &   -     \\
  8 &  5.35 &  4.81 &   7.20 &  21.66 & 22.73 \\
  9 &  4.49 &  5.78 &   7.32 &  22.23 & 23.48   \\
 10 &  8.66 & -4.07 &   9.57 &  25.18 &   -     \\
 11 & -4.75 &  8.78 &   9.99 &  22.83 & 24.65   \\
 12 & 11.34 & -2.19 &  11.55 &  24.58 &   -     \\
 13 &  0.87 &-11.70 &  11.73 &  25.28 &   -     \\
 14 & -4.54 &-11.63 &  12.49 &  25.06 &   -     \\
 15 & 12.55 & -0.40 &  12.56 &  23.58 &   -     \\
 16 & 10.47 &  8.00 &  13.18 &  24.54 &   -    \\
 17 &-13.44 &  6.95 &  15.13 &  24.73 &   -    \\
 18 & -7.32 & 13.29 &  15.17 &  24.96 &         \\
 19 & 14.96 &  3.91 &  15.46 &  25.31 &   -     \\
 20 & 11.20 & 10.68 &  15.48 &  23.99 & 25.41   \\
 21 & -5.17 & 15.36 &  16.20 &  24.63 &   -     \\
 22 & -3.63 &-16.08 &  16.49 &  24.04 &   -    \\
 23 & 10.99 & 12.52 &  16.66 &  24.25 &   -     \\
 24 & 11.77 & 12.43 &  17.12 &  24.09 & 25.60   \\
 25 & -7.91 &-16.59 &  18.38 &  24.44 &   -     \\
 26 & 18.33 & -4.84 &  18.96 &  24.11 & 25.55   \\
 27 & 15.36 &-12.99 &  20.12 &  25.10 &   -     \\
 28 &-18.38 &  9.04 &  20.48 &  24.35 &   -     \\
\noalign{\medskip}\hline
\end{tabular}
\end{table}
This small impact parameter together with the low ionization
level of the $\zd = 0.6295$ absorber, strongly suggest that the \lya \ line 
of this system is damped.  Boiss\'e et al. (1996) have observed this quasar
with the HST-FOS and detected a strong \lya \ absorption with $\Wr$ = 12~\AA . 
The complex line profile  does not however permit to unambiguously
conclude that this line is damped. In the assumption of a damped line, the 
\hi \ column density would reach $3\times 10^{20}$~\cm2 .

\begin{figure}[!h]
\centerline{\psfig{figure=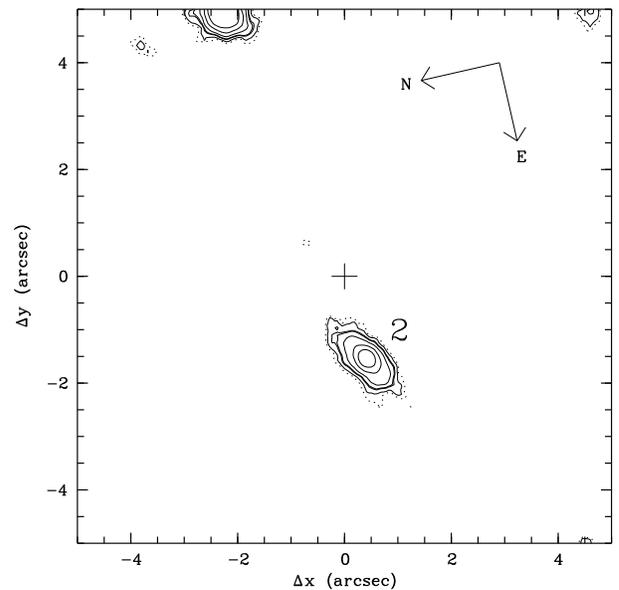,height=8cm}}
\caption[]{\label{q1209sub}Same as Fig.~\ref{ex0302sub} for Q~1209+107, with 
levels 1.5 (dotted), 2, 3, 5, 9 and 17$\sigma$}
\end{figure}

The PC2 field of Q~1209+107 is shown in Fig.~\ref{q1209field}. The galaxy
detected by Arnaud et al. (1988) is  well resolved  from the quasar image 
(object \#2). After PSF subtraction, there is no closer object to the quasar
sightline down to an impact parameter $\theta= 0.3\arcsec$ and apparent 
magnitude $\ms=25.2$ (see Fig.~\ref{q1209sub}). Consequently, we identify
this galaxy as the $\zd=0.6295$  absorber. The galaxy rest-frame magnitude 
 is $\Mb=-22.0$, and its projected 
distance to the quasar sightline  equals $D=14.5\h50$~kpc.  Its color, 
$\mq-\ms=1.5$, is compatible with that of a spiral galaxy at intermediate 
redshift (Frei \& Gunn 1994). The galaxy major and minor axes are equal to
 2.03\arcsec \ and 0.75\arcsec, leading to linear sizes of 18.4 and 
$6.8\h50$~kpc and an inclination angle of about $70\deg$.  The quasar 
sightline intersects the disk of the galaxy at a radial distance of 
$28\h50$~kpc. 

The object located 7\arcsec \ north-east to the quasar sightline,
and identified by Cristiani (1987) as the $\za=0.3930$ absorber, is resolved 
into  two well separated galaxies in our PC2 image (\#8 and \#9). Both 
galaxies are very blue, $\mq-\ms$ = 1.1 and 1.3 for objects \#8 and \#9 
respectively, which suggests an interactive pair, their projected linear 
separation 
being then equal to 10$\h50$~kpc. The elongated, bright central parts of
galaxy \#9 are embedded in a diffuse extended envelope. Recent integral 
field spectroscopic observations at CFHT by one of us (J. Bergeron) show that
the very strong optical line emission detected by Cristiani (1987) is due to 
object \#8. 

There are three possible candidates for the $\za=1.8434$ \civ \  absorber,
 all having diffuse and irregular morphologies: 
object \#4 ($D=73\h50$~kpc, $\Mb=-22.8$), which is close to a bright star
(object \#3), object \#7 ($D=90\h50$~kpc, $\Mb=-23.1$), and object \#11 
($D=129\h50$~kpc, $\Mb=-23.8$). All three candidates would then be
very luminous galaxies.
\subsection{\label{sec1229}PKS~1229$-$021 ($\ze = 1.038$)}
In the quasar radio spectrum, Brown \& \ Spencer (1979) have discovered a 
21~cm absorption system at $\zd = 0.39498$, and associated metal lines of
\mgii , \feii , \caii \ and \mgi \  have been detected by Briggs et al. 
(1985). This system may include up to 13 components spread over 250\kms 
(Lanzetta \& Bowen 1992). Another \mgii \ system  at $\za = 0.7568$ (Steidel et
al. 1994a) has an  associated Lyman Limit discontinuity discovered in the 
IUE spectrum of the quasar (Lanzetta et al. 1992), which appears roughly at
the expected wavelength for the $\zd=0.39498$ \lya \ line. Nevertheless, the
recent HST-FOS spectrum of Boiss\'e et al. (1996) clearly indicates that the
latter line is damped with  N(\hi) $\simeq 4\times 10^{20}$~cm$^{-2}$ 
(this was not {\it a priori} obvious since the radio
and optical sources do not coincide spatially). These data also reveal 
an additional \civ\ system at $\za = 0.7005$.
The 21~cm absorber has not been identified by BB91, who concluded that
it should be very close ($\theta \le 1\arcsec$) to the quasar sightline, as 
expected from the presence of 21~cm absorption. 
\begin{table}
\caption[]{\label{q1229tab}List of the objects in the field of PKS~1229$-$021}
\begin{tabular}{rrrrrr}
\hline\noalign{\smallskip}
Object & $\Delta\alpha$  & $\Delta\delta$  & $\theta$ &$\ms$ &$\mq$ \\
       & (\arcsec)       & (\arcsec)       & (\arcsec)&      &       \\
\hline\noalign{\smallskip}
  1 &  0.00 &  0.00 &  0.00 &  16.89 & 17.65 \\
  2 & -0.6  &  -0.2 &   0.6 &  24.2  & 24.4  \\
  3 &  0.3  &  -1.4 &   1.4 &  23.0  &   -   \\
  4 & -1.4  &  -0.1 &   1.4 &  25.0  &   -    \\
  5 &  1.6  &  -0.3 &   1.6 &  25.1  &   -   \\
  6 & -1.8  &  -0.1 &   1.8 &  25.8  &   -   \\
  7 & -6.76 & -3.88 &  7.80 &  24.88 &   -     \\
  8 & -7.97 &  0.69 &  8.00 &  24.61 &   -   \\
  9 & -4.61 &  6.88 &  8.28 &  22.66 &   -    \\
 10 & -4.40 &  7.86 &  9.01 &  22.11 & 23.20  \\
 11 & -3.88 & -8.79 &  9.61 &  22.89 &   -     \\
 12 & -7.90 &  7.06 & 10.59 &  24.58 &   -     \\
 13 & -5.59 & 10.35 & 11.77 &  25.31 &   -    \\
 14 & 10.32 & -7.01 & 12.47 &  23.91 & 24.37   \\
 15 & 12.42 &  2.92 & 12.76 &  25.43 &   -    \\
 16 & 11.87 &  6.68 & 13.62 &  24.93 &   -     \\
 17 & -2.06 & 13.60 & 13.75 &  22.60 &   -     \\
 18 &  0.78 &-15.82 & 15.84 &  23.57 &   -     \\
 19 & 14.88 & -6.23 & 16.13 &  24.55 &   -     \\
 20 &  8.27 & 14.53 & 16.72 &  22.73 &   -     \\
 21 &  9.55 &-14.74 & 17.56 &  25.45 &   -     \\
 22 &-12.04 & 12.94 & 17.67 &  24.79 &   -     \\
 23 &-17.95 &  1.96 & 18.06 &  24.97 &   -     \\
 24 &  5.19 & 17.89 & 18.62 &  23.12 &   -     \\
 25 & -2.99 &-19.41 & 19.64 &  23.02 & 24.33   \\
 26 &-18.00 & 10.85 & 21.02 &  24.80 &   -     \\
 27 &-19.45 &  9.19 & 21.52 &  24.30 &   -     \\
 28 &-21.09 &  7.82 & 22.49 &  23.99 &   -     \\
\noalign{\medskip}\hline
\end{tabular}
\end{table}
\begin{figure*}
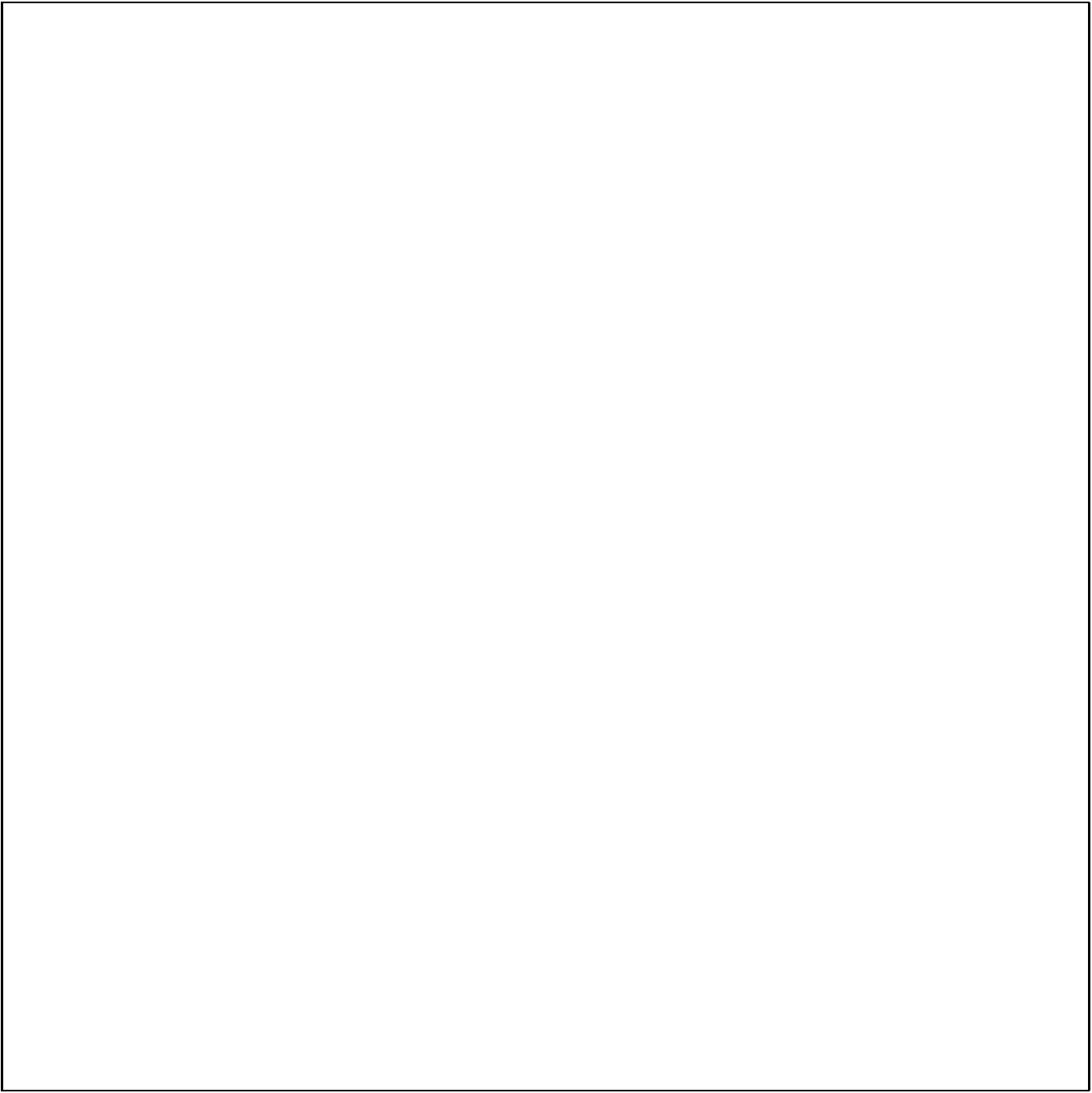

\picplace{18cm}
\caption[]{\label{q1229field}Field of the PC2 around Q~1229$-$021. Objects
are labeled as in Table~\ref{q1229tab}}
\end{figure*}
By measuring the Faraday 
rotation of the radio emission along the radio jet that extends to about
3\arcsec \ west to the quasar, and using a model of the magnetic field of a 
spiral galaxy, Kronberg et al. (1992) concluded that the observed Faraday
rotation was compatible with an intervening spiral 
galaxy being located 2\arcsec \ south-west to the quasar sightline. After PSF
subtraction performed on ground-based images of this quasar, Steidel et al. 
(1994a) have found two objects within 2\arcsec \ from the quasar image. They have 
associated the brightest southern object to the DLAS, and the eastern galaxy 
to the higher redshift absorber. 

\begin{figure}
\centerline{\psfig{figure=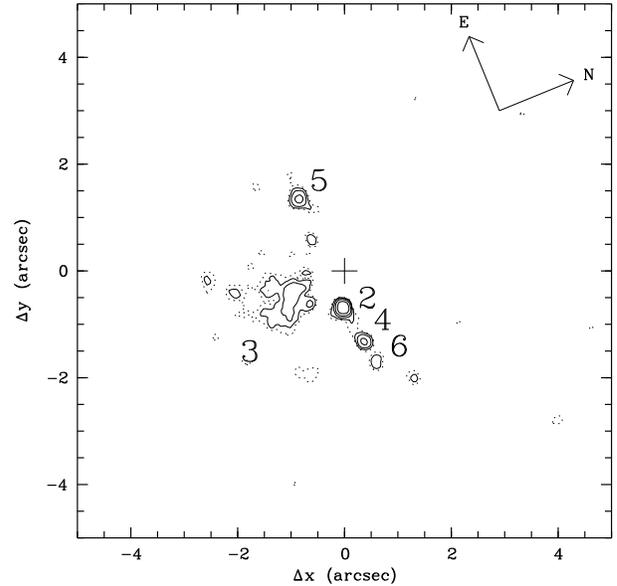,height=8cm}}
\caption[]{\label{q1229sub}Same as Fig.~\ref{ex0302sub} for PKS~1229$-$021,
  with levels 1.5 (dotted), 2, 3, 5, 9 and 17$\sigma$}
\end{figure}

The PC2 field around PKS~1229$-$021 is presented in Fig.~\ref{q1229field}.
After  PSF subtraction, we have discovered five objects very close to
the quasar sightline (see Fig.~\ref{q1229sub} and Table~\ref{q1229tab}). The 
southern object (\#3) detected in our CFHT images and by  Steidel et al. 
(1994a) is a very diffuse irregular galaxy with a peak intensity 
$\mu_{\mathrm 702, max} = 23.2$~mag~arcsec$^{-2}$. From the PC2 data, we get a
magnitude $\ms = 23.1$ and an impact parameter $\theta = 1.4\arcsec$. If 
identified as the damped \lya \ absorber, it has $\Mb=-18.9$ and  $D=9.9\h50$~kpc. 
This absorber would then 
be fairly faint. However, due to its very small pixel size, the PC2 is not very
sensitive to low surface brightness emission, and the magnitude derived from 
our CFHT  data is brighter, $\mr=22.1$ and probably more reliable. That 
measured by Steidel et al. (1994a) in the $I$-band is also high, $\mi = 21.7$. 

\begin{figure}
\centerline{\psfig{figure=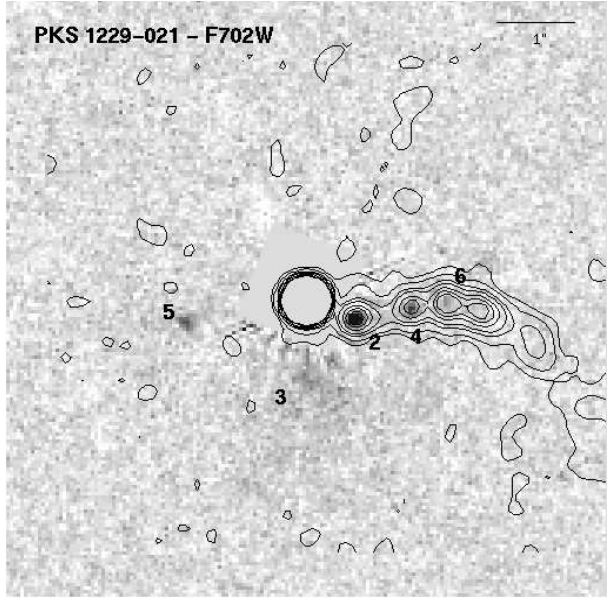,width=8cm}}
\caption[]{\label{q1229radio}$8\arcsec \times 8\arcsec$ field around 
PKS~1229$-$021, with the contours of the radio jet (Kronberg, private 
communication), which shows the alignment of objects \#2, 4 and 6 along 
the jet. North is up and east is to the left}
\end{figure}

The other four objects at small impact parameters are compact: one (\#5) is  
east to the quasar and the other three (\#2, \#4 and \#6, all
unresolved),  are well aligned. Object \#2 is also detected in the F450W 
image and is very blue with $\mq-\ms = 0.2$. This alignment draw our attention
onto a 
possible relationship with the radio jet. Indeed, the superposition of our 
F702W image on a recent higher resolution 8.16 MHz radio map that Dr. P. 
Kronberg kindly communicated to us (unpublished observations 
by Kronberg, Perley, Dyer \& Roeser), shows a remarkable correspondence 
between objects \#2, \#4, \#6 and the three first radio knots seen along 
the radio jet (see Fig.~\ref{q1229radio}).
This quasar is then one of the very few sources, and to our knowledge the 
one with the highest redshift, for which optical emission associated to a 
radio jet has been seen. It bears some resemblance to e.g. 3C~277.3 
(Miley et al. 1981; Bridle et al. 1981) and 3C~346 (Dey \& van Breugel 1994;
van Breugel et al. 1992) in the sense that optical emission arises
preferentially  
at  locations where the jet shows a break. In particular, PKS~1229$-$021 
appears unlike 3C~78 in which the spatial distribution of the optical emission 
is continuous and closely follows the unbent radio jet. Clearly, the presence 
of the three radio/optical knots is closely connected to the radio source 
morphology. At the brightest object \#2, the jet strongly interacts with some 
material and is deflected northward while, further away from the central core, 
the jet is deflected southward in a more continuous manner until it joins 
the south-west lobe. On the contrary, the (undetected) north-east flow 
expands more freely. Regarding the origin of the absorption systems, it 
therefore appears that none of objects \#2, \#4 and \#6 can be considered as 
candidate absorbers. However, the emitting gas with which 
the jet interacts may be responsible for the Faraday rotation observed by 
Kronberg et al. (1992), especially since no galaxy-like object is seen towards 
the jet. Blob \#2, which is the closest ($\theta = 0.6\arcsec$ or 
$D=6.6\h50$~kpc at $\ze = 1.038$) and the bluest object, could correspond to 
a cloud belonging to the galaxy hosting PKS~1229$-$021 on which the jet 
rebounds while the furthest 
away 
blobs  could be associated with gas entrained 
outside this galaxy by the jet. Assuming a flat spectrum, these three blobs
have absolute rest frame magnitudes $\Mb = -20.4$, $-19.6$ and $-18.7$
for \#2, \#4 and \#6 respectively.

The eastern compact blob \#5 may cause either the $\za = 0.7568$ system 
(as suggested by Steidel et al. 1994a; this \mgii \ absorber would then have a 
magnitude  $\Mb=-18.8$), or the  $\za = 0.7005$ \civ\ absorption system.  

There are 28 objects brighter than $m_{\mathrm 702, lim}=25.8$ in this 
field (Fig.~\ref{q1229field} and Table~\ref{q1229tab}). There is a $3\sigma$ 
level excess of galaxies in the magnitude range $22\le\ms\le 24$, which
may trace a group or a cluster of galaxies at the quasar 
redshift ($\ze=1.038$). This magnitude range corresponds to 
$-24.1\le\Mb\le -22.1$ at this redshift. As for 3C~196, the analysis of the 
WFC2 data will provide further information on this galaxy excess. We note
that Steidel et al. (1994a) also discovered several galaxies with optical and 
IR colors consistent with the assumption of a cluster of elliptical galaxies 
at the quasar redshift.

\begin{figure*}
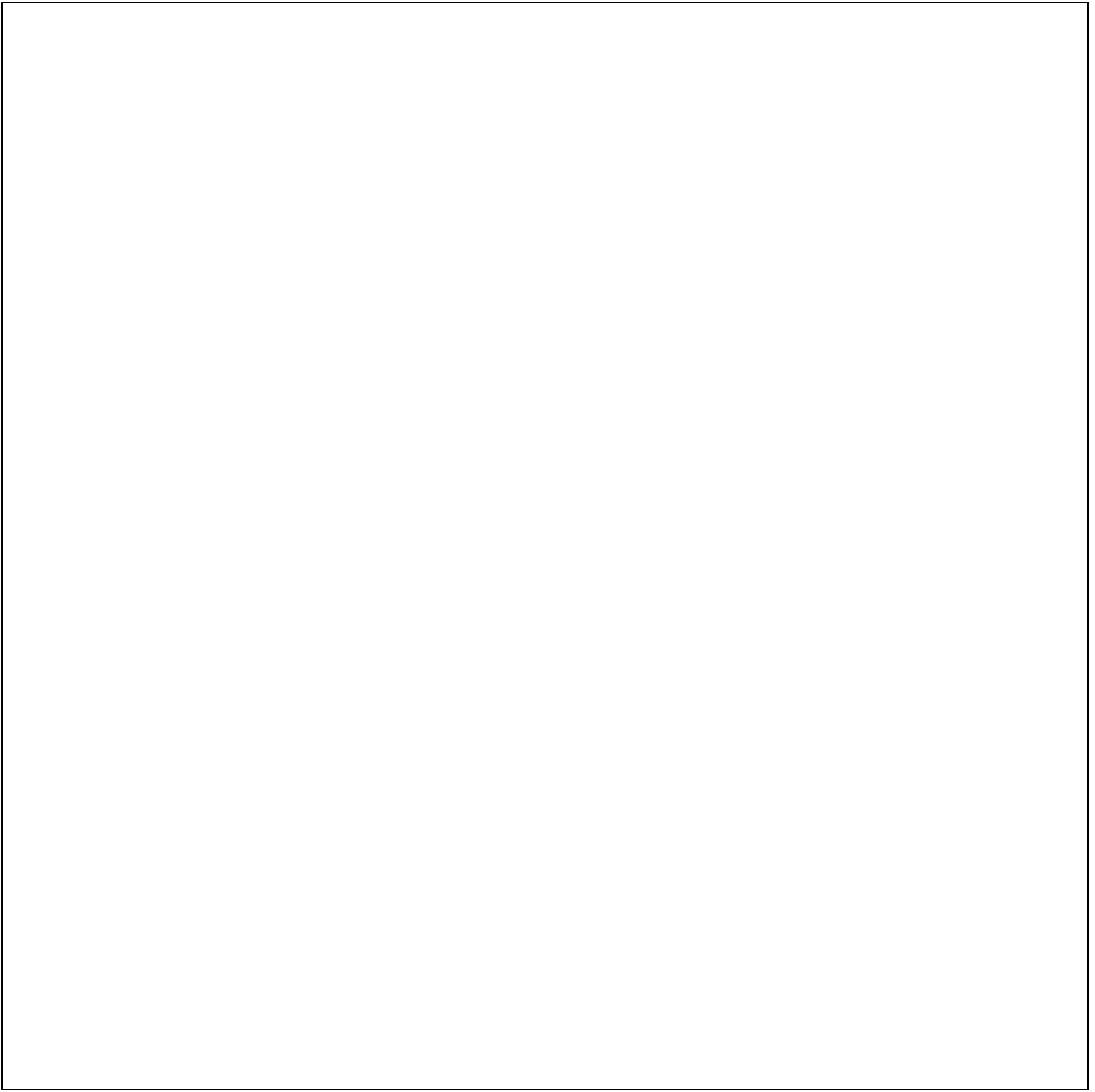

\picplace{18cm}
\caption[]{\label{q3c286field}Field of the PC2 around 3C~286. Objects
are labeled as in Table~\ref{q3c286tab}}
\end{figure*}
In ground-based images, objects \#9 and \#10 are  totally blended 
(BB91; Steidel et al. 1994a), but are clearly separated on the PC2 images. 
Object \#10 is bright and compact, and object \#9 is fainter and very diffuse. 
It is most likely that the optical emission lines detected at $z=0.199$ by
BB91 arise from the brightest object, which would then have absolute 
magnitude $\Mb=-18.5$. Its impact parameter is 
$D=37\h50$~kpc, nearly twice as large as the value predicted by the 
luminosity-halo radius scaling-law derived by BB91 for \mgii \ absorbers,  
which is consistent with the non-detection in the HST-FOS quasar spectrum  
of an associated \mgii \ absorption (Boiss\'e et al. 1996).
Object \#11 is very red, $\mq-\ms>2.3$, and could 
be  an elliptical galaxy at $z\simeq 0.2$.
\subsection{\label{sec286}3C~286 ($\ze = 0.849$)}
\begin{table}
\caption[]{\label{q3c286tab}List of the objects in the field of 3C~286}
\begin{tabular}{llllll}
\hline\noalign{\smallskip}
Object & $\Delta\alpha$ & $\Delta\delta$ & $\theta$ & $\ms$ &$\mq$ \\
       &  \arcsec       & \arcsec        & \arcsec  &       &      \\
\hline\noalign{\smallskip}
  1 &  0.00 &  0.00 &  0.00 & 16.84 & 17.99 \\
  2 &$<0.3$ &$<0.3$ &$<0.3$ &$<21.3$&       \\
  2a&  0.1  & -0.4  &  0.4  & 21.8  & 24.1  \\
  2b&  0.2  &  0.4  &  0.4  & 22.4  & 25.2  \\
  2c&  0.9  &  0.0  &  0.9  & 22.8  &   -   \\
  3 &  0.9  &  1.7  &  1.9  & 25.6  &   -   \\
  4 &  0.99 &  5.51 &  5.60 & 25.40 &   -   \\
  5 & -4.03 &  7.26 &  8.30 & 23.38 &   -   \\
  6 &  7.53 & -3.68 &  8.38 & 24.58 & 25.65 \\
  7 & -8.18 &  2.09 &  8.44 & 22.71 &   -   \\
  8 & -7.23 & -4.58 &  8.56 & 25.26 &   -   \\
  9 & -3.83 & -7.84 &  8.73 & 25.58 &   -   \\
 10 & -1.99 &  8.53 &  8.76 & 24.09 &   -   \\
 11 & -8.87 &  1.49 &  9.00 & 25.23 &   -   \\
 12 & -7.28 &  6.25 &  9.60 & 24.49 &   -   \\
 13 &  9.47 & -3.03 &  9.95 & 24.66 &   -   \\
 14 & 10.72 &  1.79 & 10.87 & 21.01 & 24.45 \\
 15 & -4.15 & 10.19 & 11.00 & 24.58 &   -   \\
 16 & 11.13 & -1.98 & 11.30 & 25.79 &   -   \\
 17 & -7.11 &-10.21 & 12.44 & 25.75 &   -   \\
 18 &-14.02 &  0.06 & 14.02 & 23.61 & 25.45 \\
 19 &-14.15 &  3.40 & 14.55 & 25.06 &   -   \\
 20 & -5.58 & 13.80 & 14.89 & 23.28 & 24.59 \\
 21 &-14.90 & -1.97 & 15.03 & 25.70 &   -   \\
 22 & -5.13 & 14.31 & 15.20 & 25.83 &   -   \\
 23 & 10.85 &-11.30 & 15.67 & 25.36 & 24.49 \\
 24 & -6.07 & 14.67 & 15.88 & 25.73 &   -   \\
 25 & -2.13 & 16.15 & 16.29 & 25.31 &   -   \\
 26 & -6.75 & 14.88 & 16.34 & 25.73 &   -   \\
 27 & 15.40 & -7.21 & 17.01 & 25.42 &   -   \\
 28 & -2.47 & 16.86 & 17.04 & 24.92 & 24.45 \\
 29 & 17.14 & -4.23 & 17.65 & 24.68 &   -   \\
\noalign{\medskip}\hline
\end{tabular}
\end{table}
A 21~cm absorption line was detected at $\zd = 0.692$ by Brown \&
Roberts (1973), and shows a single, narrow ($b=5$\kms ) component. The 
associated \mgii \ and \feii \ lines have been observed by Spinrad \& McKee 
(1979). Using Pre-COSTAR FOS observations, Cohen et al. (1994), have shown 
that the \lya \ line is damped, and their estimate of the \hi \ column 
density is 
$\simeq 2\times 10^{21}$\cm2. Using this value and their own measurements of 
\feii, 
\znii \ and \crii, Meyer \& York (1992) derived a very low metal abundances of 
$Z_\odot / 17$, for a look back time comparable to the age of our solar system.
Steidel et al (1994a) have obtained ground-based broad-band images of this 
quasar, 
and after PSF subtraction, they detected a galaxy of low surface brightness, 
$\mu(I_{\mathrm  AB})=25.5$~mag~arcsec$^{-2}$,  2.5\arcsec \ away from 
the quasar sightline. If at the absorption redshift, its 
linear impact parameter would be $D=24\h50$~kpc. Furthermore, Steidel et
al. (1994a), claimed that this low surface brightness could explain the low
metallicity of the absorber since (according to e.g Mc Gaugh 1994),
low-surface brightness galaxies have a slower chemical evolution than 
``normal'' galaxies as the Milky Way.

\begin{figure}
\centerline{\psfig{figure=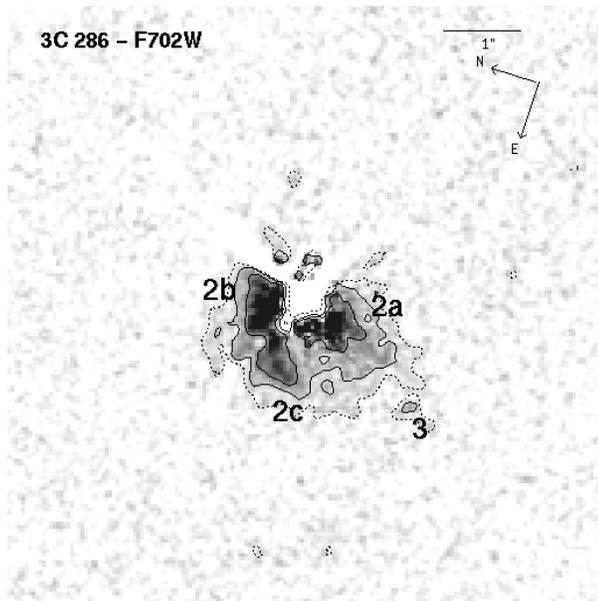,height=8cm,clip=t}}
\caption[]{\label{q3c286sub}$10\arcsec\times10\arcsec$ field around 3C~286,
  showing the objects as labeled in Table~\ref{q3c286tab}. The $1.5\sigma$
  (dotted) $3\sigma$ and $7\sigma$ level contours have been overplotted}
\end{figure}

In Fig.~\ref{q3c286field}, we show the PC2 field around 3C~286. 
At the location of the object detected by Steidel et al. (1994a), 2.5\arcsec \
south-east to the quasar sightline, we do not  detect any extended emission, 
most probably because PC2 observations are not as sensitive to low surface 
brightness emission as ground-based images. However, south-east to the quasar 
image, aside from a very faint  object (object \#3 in Fig.~\ref{q3c286sub}), 
with $\ms=25.6$ and an impact parameter of 1.9\arcsec \, we detect 
diffuse emission that we
identify with the inner part of Steidel et al.' amorphous object.
After  quasar profile subtraction, we do find a very bright object roughly
centered on the quasar (see Fig.~\ref{q3c286sub}). There are two main 
components along the south-north axis (labeled \#2a and \#2b), and  a bright 
extension of lower surface brightness at the south-east 
(\#2c). It is  possible that all these components belong to the damped \lya \ 
absorber, and its magnitude would then be  $\Mb < -22.0$.
A possible alternative is that objects \#2a and \#2b, which are located
roughly symmetrically to the quasar, are both  part
of the quasar's host galaxy, which central region lies on the saturated part
of the quasar image, while object \#2c is  the damped \lya \ absorber. 
There is some spatial overlap between the  object found by Steidel et al. 
(1994a) and the diffuse emission detected in our PC2 image south of object \#2c.
 The latter could also have a western 
extension blended with  the quasar image residuals. This would be required to 
account for the 21 cm absorption, since   more than 90\% of the radio 
emission arises from the core of 
the source (Spencer et al. 1991 and references therein). Identifying  object 
\#2c as  the damped \lya/21 cm absorber leads to a   magnitude $\Mb=-20.5$ 
and  an impact parameter $D=8.5\h50$~kpc. The magnitude of the quasar's  
host galaxy would then be brighter than $\ms = 21.8$, thus $\Mb<-22.3$. 

There are 29 objects brighter than $m_{\mathrm 702, lim}=26.2$ in the PC2 
field around 3C~286 (Fig.~\ref{q3c286field}). The most peculiar is object \#14, 
which has a very elongated disturbed morphology, with a length of 3\arcsec \ 
and 
a width of 0.71\arcsec, a peak surface brightness 
$\mu_{\mathrm 702, max}=22.3$~mag~arcsec$^{-2}$, and a very red color,
$\mq-\ms=3.4$. This object shows three emission lines 
([O\,{\sc ii}]$\lambda$3727 and [O\,{\sc iii}]$\lambda$$\lambda$4959,5007)   
at $z=0.3338$, which leads to 
$D=68\h50$~kpc and $\Mb = -20.0$ (Le Brun et al. 1997).
\subsection{\label{sec1331}MC~1331+170 ($\ze = 2.084$)}
There are four certain, metal-rich absorption systems detected in the spectrum of 
this quasar, of which two are multiple. The lowest redshift system is double, 
$\zd=0.7443, 0.7454$, which corresponds to a velocity separation of  
170\kms;  both components show  \mgii \ and \feii \ 
 absorption lines (Sargent et al. 1988),  indicative of a DLAS.  In the high 
spectral resolution data of Churchill et al. (1995), there are in fact six 
components, including the two quoted above, spread over 470\kms, and 
symmetrically disposed. This is typical of a radial inflow/outflow (Churchill 
et al. 1995), or  a disk seen edge-on. The existence of a strong DLAS at 
higher redshift prevents observation of the \lya \ line from this system.
\begin{figure*}
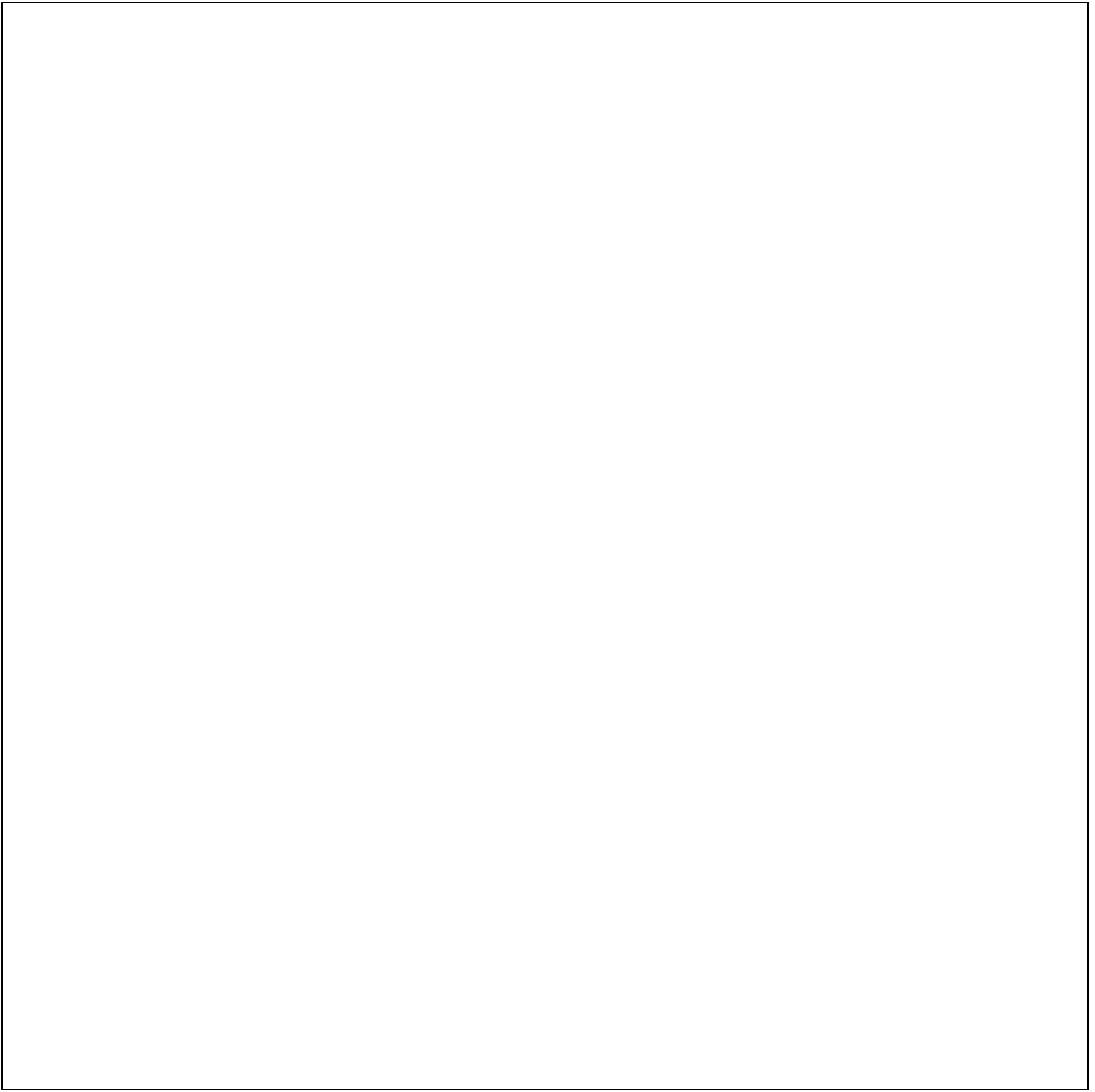

\picplace{18cm}
\caption[]{\label{mc1331field}Field of the PC2 around MC~1331+170. Objects
are labeled as in Table~\ref{mc1331tab}.}
\end{figure*}
There is a strong \mgii \ doublet at $\za = 1.3284$, blended with the 
\feii$\lambda 2344$ lines at $\za = 1.776, 1.786$ (Steidel \& Sargent 1992). 
Sargent et al. (1988) detected a \civ \ doublet at $\za =1.4462$, too faint 
to be included in the sample of Young et al. (1982), and these are the only 
lines seen at this redshift. The $\za=1.786$ component of the high-redshift 
double system shows strong \civ\ 
and \mgii\ absorption, whereas that at  $\zd=1.776$ has been detected 
in 21 cm (Wolfe \& Davis 1979) and has an associated damped 
\lya \ absorption from which Chaffee et al. (1988)  derived an \hi \ column 
density of $1.5\times 10^{21}$\cm2. There are  32 lines 
associated with this system, from C\,{\sc i} and O\,{\sc i}
to \civ \ and Si\,{\sc iv}, and the derived carbon abundance is similar to 
the Galactic value. The   $\za=1.786$ system exhibits  \mgii \ and 
\feii \ absorption, but no associated 21 cm absorption.

High spatial resolution images taken at CFHT (Churchill et al. 1995; Bergeron 
unpublished) show several faint objects around the quasar sightline. Among these, 
there is an edge-on spiral galaxy 2\arcsec \ from the quasar sightline and 
with its major axis nearly aligned with the quasar sightline. This favorable 
geometry, the 
brightness of the galaxy and the velocity splitting of the lower redshift DLAS 
candidate led us to tentatively identify this object as the $\zd = 0.7443$ 
absorber prior to the results of this HST survey. No redshift is 
available for any of these objects. 

\begin{figure}
\centerline{\psfig{figure=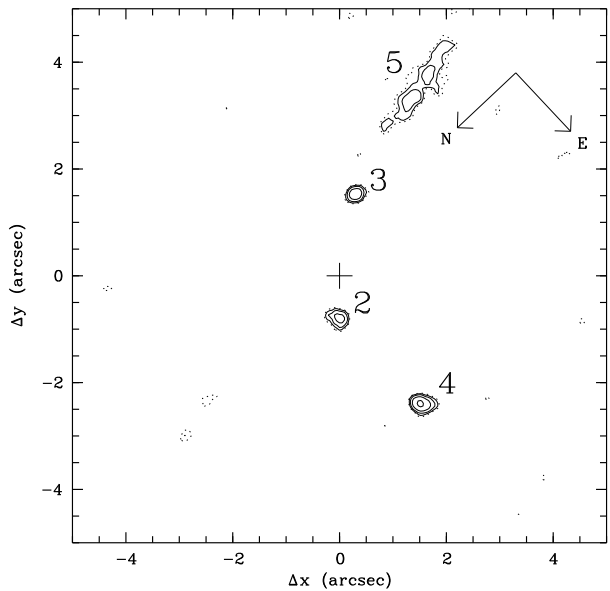,height=8cm}}
\caption[]{\label{mc1331sub}Same as Fig.~\ref{ex0302sub} for MC~1331+170,
  with levels 1.5 (dotted), 2, 3, 5 and 9$\sigma$}
\end{figure}

Since there is a confirmed  DLAS at high redshift, images were 
taken in the  F702W and  F814W filters (the latter corresponds roughly to the 
standard $I$ filter). The F702W filter image of the PC2 is presented in
Fig.~\ref{mc1331field}. After PSF subtraction, the HST images reveal one faint 
object 
very close to the quasar sightline (object \#2 in Fig.~\ref{mc1331sub}), 
at an impact parameter $\theta = 0.75\arcsec$ and with a magnitude $\ms=24.9$. 
The other objects (\#3 to \#5) that appear on Fig.~\ref{mc1331sub} were 
already visible before PSF subtraction (Fig.~\ref{mc1331field}). 
 We have also removed the quasar image on the F814W frame using the PSF 
derived for the F702W images. Even if the shape of the PSF is color 
dependent, the resulting image  is adequate for object detection at  impact
parameters $\theta \geq$ 0.6\arcsec.  Object \#2 is clearly  detected in the 
F814W image, but the estimate of its magnitude is uncertain due to 
important PSF residuals from the quasar image. 
\begin{table}
\caption[]{\label{mc1331tab}List of the objects in the field of MC~1331+170}
\begin{tabular}{rrrrll}
\hline\noalign{\smallskip}
Object & $\Delta\alpha$  & $\Delta\delta$  & $\theta$ &$\ms$ &$\mh$ \\
       & (\arcsec)       & (\arcsec)       & (\arcsec)&      &       \\
\hline\noalign{\smallskip}
  1 &  0.00 & -0.00 &  0.00 & 16.91 & 16.55 \\
  2 &  0.58 &  0.48 &  0.75 & 24.9  & 23.8$^{\mathrm a}$\\
  3 & -0.90 & -1.30 &  1.58 & 25.1  & 24.4   \\
  4 &  2.82 &  0.50 &  2.86 & 24.2  & 24.2   \\
  5 & -1.53 & -3.54 &  3.86 & 21.40 & 21.47  \\
  6 &  2.53 &  6.61 &  7.08 & 25.50 &   -    \\
  7 & -5.27 &  5.21 &  7.41 & 25.44 & 25.12  \\
  8 & -2.13 & -7.54 &  7.84 & 25.47 & 24.50  \\
  9 & -6.35 & -4.96 &  8.06 & 24.37 & 23.82  \\
 10 &  7.79 & -3.23 &  8.43 & 25.73 & 24.36  \\
 11 & -6.39 &  7.88 & 10.14 & 25.06 & 25.29  \\
 12 &-10.99 & -0.52 & 11.00 & 24.94 & 24.92  \\
 13 &-10.47 & -4.87 & 11.55 & 24.96 & 24.66  \\
 14 & -3.80 &-11.17 & 11.80 & 24.21 & 24.17  \\
 15 &  6.39 & 10.79 & 12.54 & 25.57 &   -    \\
 16 & -7.21 &-10.82 & 13.01 & 25.62 &   -    \\
 17 &  0.19 &-13.38 & 13.38 & 25.71 &   -    \\
 18 & 13.94 & -7.25 & 15.71 & 25.61 &   -    \\
 19 &-11.63 &-11.02 & 16.02 & 25.82 & 24.60  \\
 20 & 15.81 &  5.07 & 16.60 & 25.49 & 25.11  \\
 21 &-15.41 & -7.80 & 17.27 & 24.76 & 23.99  \\
 22 & 18.21 &  2.28 & 18.36 & 25.02 & 24.58  \\
 23 &  2.54 & 18.37 & 18.55 & 25.10 & 25.16  \\
 24 & 19.33 & -0.02 & 19.33 & 24.76 & 24.50  \\
 25 & 20.04 & -1.15 & 20.08 & 24.36 & 24.22  \\
 26 &  2.45 &-20.57 & 20.71 & 25.86 &   -    \\
\noalign{\medskip}\hline
\end{tabular}
\smallskip\par
$^{\mathrm a}$ The error on the magnitude of this object may be large, since 
it is partly located in a region where subtraction residuals are present\\
\end{table}
The two closest objects (\#2 and \#3) could be at $z\simeq 1.78$, may be part 
of a group, and give rise to the $\zd=1.776$ and $\za=1.786$ absorption systems.  
They would then have impact parameters  $D = 9.5\h50$~kpc and $20.0\h50$~kpc 
and magnitudes $\Mb=-23.0$ and $-22.7$ respectively (note however that the
$k$-correction needed to estimate the rest-frame $B$ magnitude is quite large 
at this redshift even for the F814W data (1.7 mag), and that the uncertainty 
is about 0.5 mag). 
Object \#4 would be intrinsically very luminous, unless it were at 
$\zd = 0.7443$. 
If it is identified as the $\za=1.3284$ absorber, its magnitude and impact 
parameter would be $\Mb = -23.6, D=34\h50$~kpc, i.e. within the range found for 
lower redshift \mgii \ absorbers, although on the bright end.
The $\za = 1.4462$ \civ \ absorption could be produced by  any one  of the 
objects \#6 to \#10. Their absolute magnitudes  and impact parameters would 
then be within the ranges $ -22.0 \leq \Mb \leq -20.6$ and $86 \leq D \leq  
100\h50$~kpc. 

The nearly edge-on spiral galaxy (object \#5) is clearly visible, 3.86\arcsec \ 
south to  
the quasar sightline. Its color is $\ms-\mh=-0.1$, which indicates a flat
spectrum. It is the most likely candidate for the $\zd = 0.7443$ absorber, with 
an impact parameter of $38\h50$~kpc and an absolute magnitude $\Mb = -22.9$. 
Assuming that this galaxy is strictly edge-on, then the absorbing clouds 
are located $9\h50$~kpc above the galactic plane, at a projected distance of 
$36\h50$~kpc from the galactic center.
\section{\label{disc}Discussion}
At $z\sim$ 2-3, the 21cm/damped \lya \ absorption-selected galaxies are 
assumed to trace protogalactic disks, as first proposed by Wolfe et al. 
(1986). 
The aim of our survey is to ascertain whether or not the \hi \ cross-section 
selected objects belong to an homogeneous class of galaxies and to
derive the main properties of the \hi \ absorbers: size, absolute luminosity, 
morphology and color. 

At the time of the target selection, there was no DLAS known at low redshift. 
Our sample includes systems with either 21~cm absorption and/or strong \feii 
\ relative to \mgii \ absorption (Bergeron \& Stasi\'nska 1986). 
One of the strong \feii \ absorbers has been confirmed to be a damped \lya \ 
system, and the two others are candidate damped \lya \ systems, 
with $5<\Wr<10$~\AA . The sample 
comprised seven candidates or confirmed DLAS at $0.4 \le z \le 1.0$ 
and a confirmed one at $\zd = 1.776$ towards seven quasars. 

A first important result of our study is the presence of candidate absorbers 
in seven cases at projected distances $\theta < 1.6\arcsec$ from the quasar 
sightline with apparent magnitudes brighter than our $5\sigma$ threshold 
$m_{\mathrm 702, lim} \simeq 25.9$. The last case, with $\theta=3.86\arcsec$, 
is a spiral galaxy with a favorable edge-on geometry. 
The rest-frame B magnitudes of the  damped \lya \  candidate absorbers cover 
the range  $-23.0 \le M_{\mathrm B} \le -18.9$. The physical parameters of 
the candidate absorbers are listed in Table~\ref{damped}.

A most striking result is the wide variety of morphological types together 
with a large spread in luminosity: the damped \lya \ absorbers do not 
constitute an homogeneous class of galaxies. Even if for a few fields, the 
identification of the absorber requires spectroscopic confirmation
(existence of more than one possible candidates), there are unambiguous cases 
of very compact candidate absorbers. A prime example is object \#2 in the 
field of PKS~0454+039: aside from a dwarf galaxy at low redshift (object \#3
at $z=0.072$), there is 
no other object detected within 5\arcsec \ from the quasar image. This 
candidate is barely resolved and after PSF subtraction, we get a FWHM for 
its core of $1.1\h50$~kpc, with a possible underlying diffuse envelope of 
$3.8\h50$~kpc in linear extent. The sample also comprises spiral galaxies 
of normal linear sizes, as those detected in the fields of Q~1209+107 and 
MC~1331+170, and one of extremely large extent ($\sim 70\h50$~kpc) in the 
field of 3C~196. The latter two are fairly bright with k-corrected (Sbc type) 
luminosities of $L_{\mathrm B} \simeq 5.7$ and $2.7 L_{\mathrm B}^\star$. 
There 
are two case of  amorphous morphology, one of fairly high surface brightness 
(object \#3 in the field of PKS~1229$-$021), and the other one, towards 3C~286,
has a moderately bright core surrounded by a  very low surface brightness 
envelope detected only on a ground-based $I$-band image (Steidel et al. 1994a). 
The central part of this galaxy has a luminosity of the order of $L^\star$ if 
at $\zd = 0.692$. In several of the above cases, high \hi \
column density gas is thus present in regions of low stellar density.

These results show that the damped \lya \ population strongly differ from the 
\mgii \ absorption-selected galaxies, the latter being a well defined class of 
objects with homogeneous properties (BB91, Bergeron et al. 1992, 
Steidel 1993, Steidel et al. 1994b, GB). The \mgii \ absorbers have blue 
B$-$K colors and show signs of recent stellar formation activity. They are 
fairly luminous, $-24.0 \le M_{\mathrm AB}$(B)$ \le -19.4$, field galaxies 
with only one known case of a dwarf galaxy at low $z$ ($<0.1$) with  
$M_{\mathrm AB}$(B)=$-17.2$, although LMC-type \mgii \ absorption-selected 
galaxies could have been identified up to at least $z\sim 0.3$ (Le Brun et 
al. 1993, GB). These galaxies are characterized by large gaseous halos 
of typical size $D=75(L/L_{\star})^{0.3}\h50$~kpc and
abundances $[Z/H] \sim -0.1$ to $-0.3$ at $z\sim0.5$ (Bergeron et al. 1994).

In three cases, we find an excess of galaxies in the quasar field. Towards
3C~196 and PKS~1229$-$021, the quasar redshift is moderate 
($\ze$ = 0.871 and 1.038) and the detected galaxy excess could trace 
a group to which the radio-loud quasar belongs. Quasar host galaxies of high
luminosity have also been tentatively detected for 3C~196 and 3C~286. 

\begin{table}[!ht]
\caption[]{\label{damped}Characteristics of the damped \lya \ absorber 
candidates and of the other objects with smallest impact parameter}
\begin{tabular}{llllll}
\hline\noalign{\smallskip}
$\zd$ or $\za^{\mathrm a}$&Obj.&$D^{\mathrm b}$&$\Mb^{\mathrm c}$&
size$^{\mathrm b,d}$&Comments \\
\hline\noalign{\smallskip}
\multicolumn{6}{c}{0302$-$223 field : $\ze = 1.400,~\zd = 1.0095$}\\
&&&&&\\
{\bf 1.0095} &2& 12.0 &$-$20.4 & $4.0 \times  3.0$ &Semi compact\\
             &3& 27.4 &$-$22.0 & $1   \times    2$ &Compact\\
             &7& 84.3 &$-$23.4 & $7.0 \times  6.4$&$\zg = 1.0095$(1)\\
\hline\noalign{\smallskip}
\multicolumn{6}{c}{0454+039 field : $\ze = 1.345,~\zd = 0.8596$}\\
&&&&&\\
{\bf 0.8596} &2&  8.3 &$-$20.5 & $1   \times    1$ &  Compact (2)\\
\hline\noalign{\smallskip}
\multicolumn{6}{c}{0809+483 field : $\ze = 0.871,~\zd = 0.437$ }\\
&&&&&\\
{\bf 0.437}  &4&   12.5 &$-$22.1   &$67 \times 19$ & Giant Sbc (3)\\
             &4&   (Bar)&$-$21.6   &                   &               \\
0.871        &2&$<$3.1& $<$$-23.1$ & $25 \times 20$& QSO host gal.\\
             &3& 11.5 &$-$22.0  &$20 \times 10$ & QSO Comp.\\
\hline\noalign{\smallskip}
\multicolumn{6}{c}{1209+107 field : $\ze = 2.191,~\zd = 0.6295$}\\
&&&&&\\
{\bf 0.6295} &2& 14.6 &$-$22.0 &$17.1 \times  7.1$ & Spiral Gal.\\
\hline\noalign{\smallskip}
\multicolumn{6}{c}{1229$-$021 field : $\ze = 1.038,~\zd = 0.39498$}\\
&&&&&\\
{\bf 0.39498} &3&  9.9 &$-$18.9 & $8.8 \times  9.0$ & LSB\\
1.038        &2&  6.6 &$-$20.8 & Unres.        & Radio knot\\
             &4& 15.5 &$-$19.6 & Unres.        & Radio knot\\
             &6& 19.9 &$-$18.7 & Unres.        & Radio knot\\
\hline\noalign{\smallskip}
\multicolumn{6}{c}{1328+307 field : $\ze = 0.849,~\zd = 0.692$}\\
&&&&&\\
{\bf 0.692} &2c&  8.5 & -20.5 & $8.7 \times  5.3$ & DLAS cand. \\
0.849    &$\begin{array}{c}2a\\2b\end{array}$& $<$$2.8$ & $<$$-22.3$&
$27 \times 11$ & QSO host gal.? \\
\hline\noalign{\smallskip}
\multicolumn{6}{c}{1331+170 field : $\ze = 2.084,~\zd = 0.7443, 1.776$}\\
&&&&&\\
{\bf 0.7443}  &5& 37.7 &$-22.9$ & $22.5 \times 3.9$&Edge-on Spiral\\
&&&&&\\
{\bf 1.776}  &2&  9.5 &$-23.0^{\mathrm e,f}$& Unres. & \\
             &3& 20.0 &$-22.7^{\mathrm e}$  & Unres.$\times 1$ & Compact\\
\noalign{\medskip}\hline
\end{tabular}
\smallskip\par
$^{\mathrm a}$ The redshifts of the  DLAS and  DLAS candidates are written in bold
characters\\
$^{\mathrm b}$ All impact parameters are given in unit of $\h50$~kpc\\
$^{\mathrm c}$ The absolute magnitudes have been calculated assuming $\zg = 
\zd$ (or $\za$), with $k$-correction (see text for details)\\
$^{\mathrm d}$ For objects labeled as ``compact'', the size is calculated
from the deconvolved angular size of the object, $\sqrt{fwhm_{\mathrm obj}^2-
fwhm_{\mathrm psf}^2}$. For well resolved objects, the extent of the $1.5\sigma$ 
isophote is given\\
$^{\mathrm e}$ These magnitudes are derived from the F814W images\\
$^{\mathrm f}$ This magnitude may be strongly affected by the subtraction 
residuals\\
References : 1 - GB, 2 - Steidel et 
al. (1995), 3 - Cohen et al. (1996)\\
\end{table}
In our study, we find several cases of galaxies very close on the sky to 
background quasars and with redshifts of about 0.5 (if these galaxies are 
assumed to be the damped \lya \ absorbers). This is an ideal situation in 
which multiple images of the quasar induced by gravitational deflection may 
be expected. Further, since we have deep and high-spatial resolution frames, 
secondary images of the QSOs induced by the absorber could be found easily, 
even if the angular separation to the primary image and  the image flux 
ratio are small. However, in none of the investigated fields do we find any 
evidence for multiple images. In some cases, a faint secondary image may be 
difficult to find if it appeared superimposed onto the image of the 
intervening galaxy (3C~196, 3C~286). For PKS~1229$-$021, object \#2 could have 
been a secondary image but, it is much more likely to be intrinsic emission 
from the jet, as indicated by the remarkable coincidence between the optical 
and radio knots (see Sect.~\ref{sec1229}). Around PKS~0454+039 and 
Q~1209+107 for instance, we estimate that we can rule out the presence of 
an unresolved object at an angular distance exceeding 0.4\arcsec \ from 
the quasar and with a flux ratio to the primary image greater than about 0.001.
Among the confirmed gravitational lenses (see e.g. Refsdal \& Surdej 1994), 
angular separations and flux ratio well above these limits have been observed. 
This already tells us that the mass of the dark halo in these lensing
galaxies is much larger than that of the ordinary ``absorption-selected'' 
galaxies discussed in this study. Our negative result is also consistent 
with the fact that in gravitational lenses, several galaxies are often at 
work to produce the multiple images. We thus note that our data 
could be 
used to set tight constraints on the mass and/or mass distribution of the 
absorbing galaxies. This would require a detailed analysis of each individual
case, which is beyond the scope of this study.

\acknowledgements{We are very grateful to Dr. P. Kronberg for providing us 
with its high resolution radio data of PKS~1229$-$021. We also would like to 
thank R. Hook (ST-ECF), P. M\o ller (STScI) and M. Giavalisco (STScI) for 
very useful advice concerning profile subtraction, and B. Fort and J. Roland
for very helpful discussions about multiple gravitational images and optical
emission associated with radio sources. We also thank the referee, M. Dickinson,
for very useful remarks}


\begin{thebibliography}{}
\bibitem[]{} Arnaud J., Hammer F., Jones J., Le Fevre O., 1988, A\&A, 206, L5
\bibitem[]{} Bergeron J., Boiss\'e P., 1991, A\&A 243, 344 (BB91)
\bibitem[]{} Bergeron J., Cristiani S., Shaver P., 1992, A\&A 257, 417
\bibitem[]{} Bergeron J., Petitjean P., Sargent W.L.W., et al., 1994, ApJ
  436, 33 
\bibitem[]{} Bergeron J., Stasi\'nska G., 1986, A\&A 169, 1
\bibitem[]{} Bertin E., Arnouts S., 1996, A\&A, in press
\bibitem[]{} Boiss\'e P., Bergeron J., Le Brun V., Deharveng J.-M., 1996, in 
  preparation
\bibitem[]{} Boiss\'e P., Boulade O., 1990, A\&A, 236, 291
\bibitem[]{} Briggs F.H., Turnshek D.A., Schaeffer J., Wolfe A.M., 1985, ApJ 
  293, 387
\bibitem[]{} Bridle A.M., Fomalont E.B., Patimaka J.J., Willis A.G., 1981, 
  ApJ 248, 499 
\bibitem[]{} Brown R.L., Mitchell K.J., 1983, ApJ 264, 87
\bibitem[]{} Brown R.L., Roberts M.S., 1973, ApJ 184, L7
\bibitem[]{} Brown R.L., Spencer R.E., 1979, ApJ 230, L1
\bibitem[]{} Brown R.L., Broderick J.J., Johnston K.J., 1988, ApJ 329, 138
\bibitem[]{} Carilli C.L., van Gorkom J.H., 1989, Nature 338, 134
\bibitem[]{} Chaffee F.H. Jr, Black J.H., Foltz C.B., 1988, ApJ 335, 584
\bibitem[]{} Coleman G.D., Wu C.-C., Weedman D.W., 1980, ApJS 43, 393
\bibitem[]{} Churchill C.W., Vogt S.S., Steidel C.C., 1995, From metal-line 
  absorption profiles to halo kinematics. In: Meylan G. (ed.) QSO absorption 
  lines, Springer, Berlin, p. 153 
\bibitem[]{} Cohen R.D., Barlow T.A., Beaver E.A., et al., 1994, ApJ 421, 453
\bibitem[]{} Cohen R.D., Beaver E.A., Diplas A., et al., 1996, ApJ 456, 132  
\bibitem[]{} Cristiani S., 1987, A\&A 175, L1
\bibitem[]{} Dey A., van Breugel W.J.M., 1994, AJ 107, 1977
\bibitem[]{} Foltz C.B., Chaffee F.H., Wolfe A.M., 1988, ApJ 335, 35
\bibitem[]{} Frei Z., Gunn J.E., 1994, AJ 108, 1476
\bibitem[]{} Guillemin P., Bergeron J., 1996, A\&A, submitted (GB)
\bibitem[]{} Holtzman J.A., Burrows C.J., Casertano S., et al., 1995, PASP 
  107, 1065
\bibitem[]{} Koratkar A.P., Kinney A. L., Bohlin R.C, 1992, ApJ 400, 435
\bibitem[]{} Kronberg P.P., Perry J.J., Zukowski E.L., 1992, ApJ 387, 528
\bibitem[]{} Lanzetta K.M., Bowen D.V., 1992, ApJ 391, 48
\bibitem[]{} Lanzetta K.M., Turnshek D.A., Sandoval J., 1992, ApJS 84,109
\bibitem[]{} Lanzetta K.M., Wolfe A.M., Turnshek D.A., 1995, ApJ 440, 435
\bibitem[]{} Le Brun V., Bergeron J., Boiss\'e P., Christian C., 1993, A\&A 
  279, 31
\bibitem[]{} Le Brun V., Bergeron J., Boiss\'e P., 1997, in preparation
\bibitem[]{} Oren A.L., Wolfe A.M., 1995, ApJ 445, 6240
\bibitem[]{} Mc Gaugh S.S., 1994, ApJ 426, 135
\bibitem[]{} Miley G.K., Heckman T.M., Butcher H.R., van Breugel W.J.M., 1981,
  ApJ 247,L5 
\bibitem[]{} Meyer D.M., York D.G., 1992, ApJ 399, L121 
\bibitem[]{} Petitjean P., Bergeron J., 1990, A\&A 231, 309
\bibitem[]{} Pettini M., Smith L.J., Hunstead R.W., King D.L., 1994, ApJ 426, 79
\bibitem[]{} Refsdal S., Surdej J., 1994, Rep Prog Phys 56, 117
\bibitem[]{} Sargent W.L.W., Boksenberg A., Steidel C.C., 1988, ApJS 68, 539
\bibitem[]{} Spencer R.E., Schilizzi R.T., Fanti C., et al., 1991, MNRAS 250,
  225 
\bibitem[]{} Spinrad H., McKee C., 1979, ApJ 232, 54
\bibitem[]{} Steidel C.C., 1993, The properties of absorption line selected
  high redshift galaxies. In Schull J.M., Thronson H. (eds.) The evolution of
  galaxies and their environment, Proceedings of the Third Grand Teton Summer
  Astrophysics Conference, Kluwer, Dortrecht 
\bibitem[]{} Steidel C.C., Dickinson M., Bowen D.V., 1993, ApJ 413, L77
\bibitem[]{} Steidel C.C., Pettini M., Dickinson M., Persson S.E., 1994a, 
  AJ 108, 2046 (SPDP)
\bibitem[]{} Steidel C.C., Bowen D.V., Blades J.C., Dickinson M., 1995, ApJ 
  440, L45 
\bibitem[]{} Steidel C.C., Dickinson M., Persson S.E., 1994b, ApJ 437, L75
\bibitem[]{} Steidel C.C., Sargent W.L.W., 1992, ApJS 80, 1
\bibitem[]{} van Breugel W.J.M., Fanti C., Fanti R., 1992, A\&A 256, 56
\bibitem[]{} Whitmore B., 1995, Photometry with the WFPC2. In: Koratkar A., 
  Leitherer C. (eds.) Calibrating Hubble Space Telescope: Post servicing 
  mission, STScI, Baltimore
\bibitem[]{} Wolfe A.M., 1987, In: Blades J.C., Turnshek D.A., Norman C.A.   
  (eds.) QSO Absorption Lines: Probing the Universe, Cambridge University
  Press, p.297
\bibitem[]{} Wolfe A.M., Davis M.M., 1979, AJ 84, 699
\bibitem[]{} Wolfe A.M., Turnshek D.A., Smith H.E., Cohen R.S., 1986, ApJS 61, 249
\bibitem[]{} Young P., Sargent W.L.W., Boksenberg A., 1982, ApJS 48, 455
\end{thebibliography}
\end{document}